\documentclass[%
reprint,
superscriptaddress,
 amsmath,amssymb,
 aps,
 pra,
floatfix,
]{revtex4-1}

\usepackage{graphicx}
\usepackage{dcolumn}
\usepackage{bm}
\usepackage{float}
\usepackage{booktabs}
\usepackage{xcolor}
\usepackage{booktabs}

\begin{document}

\preprint{APS/123-QED}

\title{Emergent epithelial elasticity governed by \\interfacial surface mechanics and substrate interaction}

\author{Ur\v ska Andren\v sek}
\email{To whom the correspondance should be addresed. \\e-mail: urska.andrensek@ijs.si}
\affiliation{Faculty of Mathematics and Physics, University of Ljubljana, Jadranska 19, SI-1000 Ljubljana, Slovenia}
\affiliation{Jo\v zef Stefan Institute, Jamova 39, SI-1000 Ljubljana, Slovenia}%

\author{Matej Krajnc}
\affiliation{Jo\v zef Stefan Institute, Jamova 39, SI-1000 Ljubljana, Slovenia}%

\date{\today}

\begin{abstract}
During the life of animals, epithelial tissues undergo extensive deformations--first to form organs during embryogensis and later to preserve integrity and function in adulthood. To what extent these deformations resemble that of non-living elastic materials is not well understood. We derive an elasticity theory of epithelia, supported by a thin layer of extracellular material and the stroma, in which the mechanics of individual cells are dominated by differential interfacial tensions stemming from cell cortical tension and adhesion. Upon coarse-graining a discrete single-cell-level mechanics model, we obtain a harmonic deformation energy and derive the critical conditions for the elastic instability, where an initially flat tissue either buckles out of plane or forms wrinkles. Due to the distinct origin of elasticity, the scaling of the critical load to induce an instability and the wrinkling wavelength with layer thickness is fundamentally different than in solid plates. The theory also naturally describes reversal of the groove-to-crest thickness-modulation phase--a recently observed epithelial shape feature which cannot be explained by the classical elasticity theory. Our work provides a guideline for understanding the relative role of cell surface tensions and the interaction of tissues with substrates during epithelial morphogenesis.

\end{abstract}

\maketitle


\section{\label{sec:level1}Introduction}

The structure and form of epithelial tissues are intricately connected to their specific functions within an organ. While the mostly protective role of the skin epithelium, for instance, requires smooth and stratified tissue architecture, epithelia of the digestive tract assume single-cell-thick corrugated structures, so as to optimize the exchange of nutrients. These distinct shapes begin to emerge during embryonic development, when the organism undergoes tissue-scale shape changes, transforming initially smooth epithelia into compartmentalized and differentiated cellular structures that eventually develop into fully functional organisms \cite{Rauzi2008, collinet15, Rauzi2015}. 

While the detailed shape- and structural changes during epithelial morphogenesis are underlined by comprehensive biochemical patterning with specific spatial and temporal dynamics \cite{Hubert21, lecuit07, lecuit11, shyer17}, the more general aspects of the shape formation rely on physical mechanisms such as buckling, wrinkling, and folding. These deformation modes can be interpreted in the context of thin-plate elasticity. Whether a plate will buckle out of plane upon uniaxial compression depends on its thickness $h$. In particular, the elastic energy of the plate associated with pure compression scales with the thickness as $W_c\sim h$, whereas the elastic energy associated with buckling scales as $W_b\sim h^3$. Consequently, thin plates tend to buckle out of plane under an externally applied uniaxial load, whereas thick plates compress while remaining flat~\cite{Timoshenko1961, pocivavsek08, jiang07}. 

This interplay between bending and compression is fundamental to understanding the physical basis of epithelial morphogenesis and the resulting complex tissue architectures. Indeed, the corrugated epithelial patterns such as villi and crypts have been previously explained through a relative growth of the fast-renewing epithelium versus a mostly static stroma \cite{tallinen14,  Ackermann2021, Savin2011, tallinen16}. The area mismatch at the epithelium-stroma interface, caused by the differential growth, induces elastic stresses, leading to the formation of wrinkles with a characteristic wavelength. The wavelength is determined by a competition between the thin epithelium, which tends to buckle out of plane, and the thick stroma, which prefers pure compression \cite{cerda03, Brau2011, shyer13vili, hanezzo11, Efimenko2005}. 

\begin{figure}[H]
    \centering
    \includegraphics[width=0.95\linewidth]{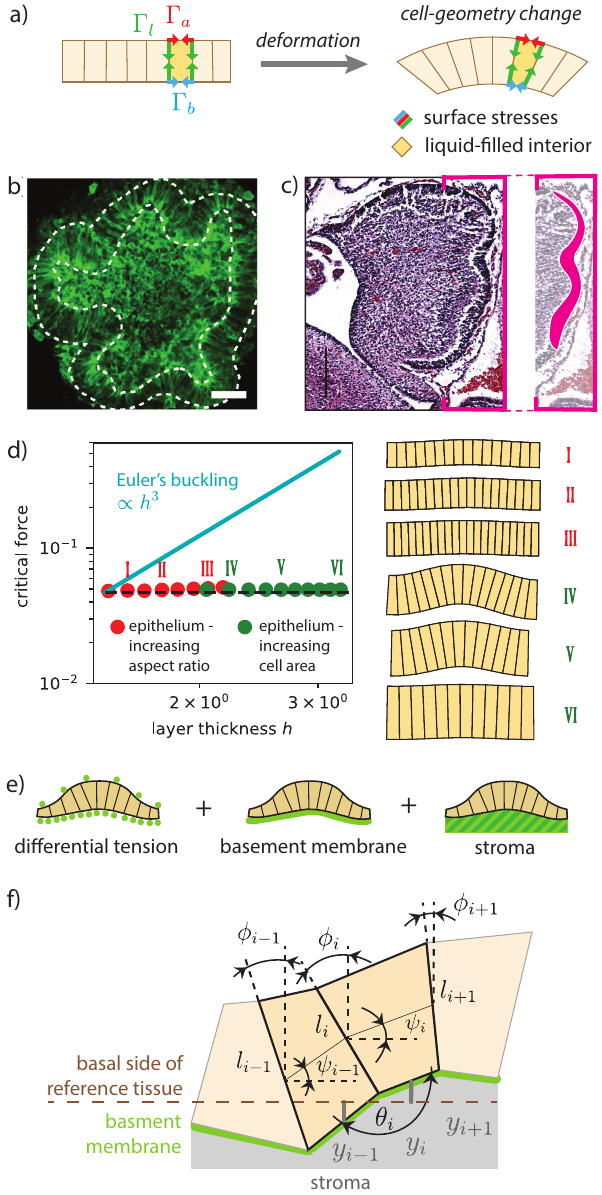}
    \caption{Vertex model of epithelial sheets with liquid/filled cells. a)~Schematic of an epithelial monolayer, composed of liquid-filled cells with surface tensions $\Gamma_a, \Gamma_b$, and $\Gamma_l$ on apical, basal, and lateral edges, respectively. Upon deformation, only surface stresses of deformed cells contribute to the elastic energy. b),~c)~+Blebbistatin organoid, reprinted from Fig. 4 \textit{A} in \cite{Karzbrun2018} (panel b) and midsaggital section of a mouse cerebellum at 17.5 embryonic days, adapted from Fig. 5 in \cite{joyner17} (panel c). Thickness modulation in both samples has an opposite phase compared to the undulation of the internal substrate. Figures in both panels are reprinted with permission from Springer Nature. Contour in panel c, marking the epithelium, was added to the original image. d)~Critical buckling force as a function of the layer's thickness. Red and green data points show the results from the vertex model, whereas the cyan line is the result for the classic Euler's buckling of a plate. Right: Shapes of uniaxially compressed epithelia for different cell aspect ratios and cell-areas. e)~Schematics of the various contributions to the total effective elastic energy of the epithelium-BM-stroma system. f)~Detailed geometry of two adjacent discrete quadrilateral cell cross-sections.}
    \label{fig:F1}
\end{figure}

However, the underlying analogy between epithelia and solid plates may disregard some cell-scale specific mechanics and thus may not always be justified. Indeed, as proposed by Thompson \cite{Thompson_1992}, the stresses in epithelia may largely be concentrated at cells' surfaces, making them behave more like liquid droplets and less like classic solid plates, where stresses are distributed throughout the volume. Within the analogy with incompressible droplets, tissue deformations contribute to an elastic energy, associated only with shape changes of cell outlines, whereas cell interiors do not support bulk elastic stresses~(Fig.~\ref{fig:F1}\textit{A}) \cite{derganc09, hannezo13}. 

This distinct origin of elasticity in epithelia may result in their specific large-scale elastic properties and deformations \cite{Fouchard20, sui2018, fernandez2021, GuhaRay24}. Indeed, as we have previously shown, epithelial layers consisting of droplet-like cells may exhibit unexpected behaviors such as forming wrinkles even when not being supported by substrates \cite{andrensek23, Rozman2021}. In addition, certain shape features of epithelia may significantly differ from the analogous shapes in solid plates. One such feature is the epithelial thickness modulation and its phase relative to the undulation of the epithelium-substrate interface. While in supported solid plates the thickness modulation is universally in phase with substrate undulations \cite{tallinen14}, an inverted phase is observed in some epithelia~(Fig.~\ref{fig:F1}\textit{B} and \textit{C}). When this phase inversion is not imposed by intrinsic substrate curvature~\cite{luciano21, Harmand22}, it cannot be explained within the classical elasticity theory~\cite{engstrom18, Karzbrun2018, joyner17} and cell surface tensions could be the root origin of the modulation \cite{krajnc15, Storgel2016, riccobelli20}.

In addition, the distinct epithelial elasticity may show up in unusual scaling relations of some of the key physical quantities such as the critical load to induce an elastic instability. This is suggested by a discrete model of uniaxially compressed epithelial monolayers, where apical, basal, and lateral cell sides are under constant surface tensions $\Gamma_a$, $\Gamma_b$, and $\Gamma_l$, respectively, while cells' interiors are assumed filled with an incompressible fluid. Using this model, we show that the critical buckling force is independent of the layer's thickness (Fig.~\ref{fig:F1}\textit{D}), in contrast to the classic Euler buckling of solid plates, where the critical force scales with thickness as $\sim h^3$.

 These suggested deviations from predictions of the classical elasticity theory highlight a need to derive elasticity theories of tissues by coarse-graining their specific microscopic mechanics models rather than {\it a priori} assuming certain macroscopic elasticity. Here, we address this challenge, by deriving an elasticity theory, describing epithelia supported by the basement membrane and the stroma~(Fig.~\ref{fig:F1}\textit{E}). To explore the role of surface stresses in shaping the tissue, we assume that cell mechanics are dominated by surface tensions, while cell interiors resemble an incompressible fluid. We find that due to these specific mechanics, the coarse-grained elasticity is fundamentally different from classic supported solid plates. We study linear stability of flat tissues and explore conditions under which epithelia wrinkle or buckle out of plane. Importantly, our theory captures the unconventional scaling of the critical in-plane force to induce an elastic instability with epithelial thickness and shows that other key physical parameters such as the wavelength of wrinkles also scale differently than in supported solid plates. Finally, our approach naturally describes the phase inversion of the groove-to-crest thickness modulation and it offers a set of experimentally testable analytical results that may help uncover the relative contribution of surface mechanics to epithelial elasticity.

\section*{Discrete cell-scale model}
We study epithelial sheets in two dimensions, where we view them as chains of quadrilateral cell cross-sections. Cell interiors are assumed incompressible, meaning that the cross-section area of each cell is kept fixed at $A_0$. Cells adhere to their immediate neighbors along the lateral sides, whereas at the basal sides they are supported by a thin layer of extracellular material, i.e., the basement membrane~(BM), and a thick stroma~(Fig.~\ref{fig:F1}\textit{F} and SI Appendix). Both BM and the stroma are considered elastic, carrying the bending- and bulk-elastic energy, respectively.

The apical, basal, and lateral cell sides are exposed to different microenvironments and carry distinct physiological functions, thereby being subjected to differential surface tensions. These tensions, denoted by $\Gamma_a$, $\Gamma_b$, and $\Gamma_l$ for apical, basal, and lateral sides, respectively, result from cell-cell- and cell-substrate adhesion, and cortical tension due to acto-myosin contractility. In discrete form, the total potential energy of the entire epithelium-BM-stroma system is thus calculated as a sum over all cells and reads
\begin{equation}
    W=\sum_i\left (\Gamma_a \ell_i^a + \Gamma_b \ell_i^b + \Gamma_l l_i + \frac{B}{2}\frac{\overline\ell^{b}_i}{\rho_i^2}  + \frac{\bar K}{2}y_i^2\ell^{b}_i\right )\>.
    \label{eq:sc_energy}
\end{equation}
Here $\ell_i^a, \ell_i^b,$ and $l_i$ are lengths of apical, basal, and lateral cell sides, respectively, $\rho_i$ is the local radius of curvature of BM, evaluated at the basal vertex of $i-$th lateral side, whereas $y_i$ is the vertical displacement of the mid-point of basal edge with index $i$ from its reference position~(Fig.~\ref{fig:F1}\textit{F} and SI Appendix). In Eq.~\ref{eq:sc_energy}, the elastic-energy densities of BM and the stroma are multiplied by the corresponding basal lengths $\overline\ell^{b}_i=(\ell^{b}_{i-1}+\ell^{b}_{i})/2$ and $\ell^{b}_{i}$, respectively; $B$ and $\bar K$ are the elastic moduli associated with BM's bending- and stroma's bulk-elastic deformations, respectively. To describe the contribution from the stroma within the linear-stability analysis, the bulk-elastic modulus $\bar K$ needs to be assumed proportional to the wavenumber of the dominant deformation mode (i.e., $\bar K=Kq$) \cite{Brau2011}.

We introduce dimensionless quantities by expressing lengths and tensions in units of $\sqrt{A_0}$ and $\Gamma_l$, respectively. In particular, we apply the following renormalizations: $l_i/\sqrt{A_0}\to l_i$ (and analogously for the other quantities with units of length), $W/(\Gamma_l\sqrt{A_0})\to W$, $B/(\Gamma_lA_0)\to B$, and $K\sqrt{A_0}/\Gamma_l\to K$, and we define dimensionless average and differential apico-basal surface tensions as
\begin{equation}
    \Gamma=\frac{\Gamma_a+\Gamma_b}{\Gamma_l}\>\>\>\>{\rm and}\>\>\>\>\Delta=\frac{\Gamma_a-\Gamma_b}{\Gamma_l}\>,
\end{equation}
respectively. Since we assume homogeneous monolayers, the values of $\Gamma$, $\Delta$, $K$, and $B$ are uniform along the tissue and, therefore, they are the only free parameters of our model.

{\bf Reference state.} In a flat configuration, all cells assume rectangular shapes and the energy~(Eq.~\ref{eq:sc_energy}) per cell simplifies to $W/N=\Gamma/h+h$, where $h$ is the cell height. Minimizing $W$ with respect to $h$ yields the optimal height
\begin{equation}
    h=\Gamma^{1/2}\>.
\end{equation}

\section*{Continuum limit}
To derive a continuum limit of $W$ (Eq.~\ref{eq:sc_energy}), we first express apical and basal edge lengths and the local radius of curvature in Eq.~\ref{eq:sc_energy} as functions of the lateral-side lengths $l_{i-1}$, $l_i$, and $l_{i+1}$, inclinations of cell midlines relative to the $x$-axis $\psi_i$ and $\psi_{i+1}$, and the inclinations of lateral sides relative to the $y$-axis $\phi_{i-1}$, $\phi_i$, and $\phi_{i+1}$~(SI Appendix). We treat these variables as well as $y_i$ and $y_{i+1}$ as functions of the distance $\sigma$ along the reference-state midline: $\phi_i \rightarrow \phi(\sigma)$ and $\phi_{i+1} \rightarrow \phi(\sigma)+ \dot{\phi}(\sigma)\sigma_0 +\ddot{\phi}(\sigma)\sigma_0^2/2$, where the "dot" denotes the derivative with respect to $\sigma$, and analogously for the other variables. 

Next, we express the energy as an integral $L=\int_0^{Nh^{-1}}\mathcal L(\sigma) {\rm d}\sigma$, where the Lagrangian density
\begin{equation}
    \mathcal L=\mathcal L_T+\mathcal L_B+\mathcal L_K+\mathcal L_{C1}+\mathcal L_{C2}
    \label{eq:Lagrangian}
\end{equation}
is expanded to second order in the variables, $\delta l(\sigma)=l(\sigma)-h$, $\psi(\sigma)$, $\phi(\sigma)$, and $y(\sigma)$, describing deformations from the reference state, characterized by $\delta l=\psi=\phi=y=0$. To simplify the equations, we hereafter omit the notation "$\delta$" from $\delta l(\sigma)$ and its derivatives. The five terms in the Lagrangian density~(Eq.~\ref{eq:Lagrangian}) are outlined below and derived in full details in~SI Appendix.

Epithelial {\bf surface energy}, captured by the first three terms of Eq.~\ref{eq:sc_energy}, in continuum limit reads
\begin{equation}
    \begin{aligned}
        \mathcal L_T&=\frac{\Delta  \ddot l \ddot  \psi }{8 h^3}+\frac{\ddot l^2}{8 h^4}+ \frac{(4 l \ddot l+\ddot \psi )^2-\ddot \psi  \ddot \phi +\ddot \phi ^2}{8 h^2 }+ \\ &+\frac{2 \Delta  \dot l \dot \psi +2\dot \psi  \ddot \psi}{4 h}+h^2  \left(\frac{1}{4} \dot l^2+\psi ^2-2 \psi  \phi+\phi ^2+4\right)+\\ &+\frac{1}{16} \left(\ddot l^2+16 \Delta  \psi \dot l+32 l^2-8 \psi  \ddot \phi -8 \dot \psi  \dot \phi -8 \phi  \ddot \psi \right)+\\ &+h \left(\dot \psi  (\psi -\phi) -\Delta  \dot \phi\right)\>.
    \end{aligned}
\end{equation}

The {\bf BM bending energy} expresses as an effective epithelial bending energy,
\begin{equation}
    \mathcal L_B=\frac{B}{2}\left (\dot\psi-\frac{\ddot l}{2}\right )^2\>,
    \label{eq:LB}
\end{equation}
with $\ddot l/2$ playing the role of a local spontaneous curvature.

The term representing the {\bf bulk elastic energy of the stroma} reads
\begin{equation}
    \mathcal L_K= {\bar K}\,\frac{\left(\ddot y+2 h \dot y+2 h  y\right)^2+4 h^4 y^2}{16 h^{4}}\>.
\end{equation}
Here, the variable $y$ is, in fact, a dependent integral quantity given by $y(\sigma)= \int _0^{\sigma} s(t) \sin{\psi(t)} {\rm d}t$, $s(t)$ being the tissue midline. By differentiating this expression with respect to $\sigma$, we obtain a constraint $\dot y = \psi  - \dot l/2$, which is taken into account by the following term in the Lagrangian density:
\begin{equation}
    \mathcal L_{C1}=Q(\sigma)h\left(\frac{\dot l}{2}-\psi +\dot y\right)\>,
    \label{eq:constraint_boundary}
\end{equation}
$Q(\sigma)$ being the corresponding Lagrange multiplier.

We study epithelial sheets under a small uniaxial compressive strain $\left |\epsilon\right |\ll 1$, imposed by a constraint $Nh^{-1}\left(1-\epsilon\right) = \int_0^{Nh^{-1}}\dot x{\rm d}\sigma$, where $x$ is the local projection of tissue midline to the horizontal axis. In the Lagrangian density, this constraint is represented by
\begin{equation}
    \mathcal L_{C2}=\mu \left(1 - \epsilon - \dot x\right)\>,
\end{equation}
where $\mu$ is the applied load ensuring a fixed compressive strain $\epsilon$. The variable $\dot x(\sigma)$ is derived from a local constraint of fixed cell-cross-section area~(SI Appendix).

\section*{Deformation modes}
The Lagrangian yields a system of three differential equations for variables $\psi(\sigma)$, $\phi(\sigma)$, and $l(\sigma)$~(SI Appendix). We solve it in the Fourier space by assuming $\psi(\sigma)=\int \Psi(q) \exp({\rm i} q \sigma){\rm d} q$,   $\phi(\sigma)=\int\Phi(q) \exp({\rm i} q \sigma){\rm d} q$, and $l (\sigma)=\delta h+\int\Lambda(q)\exp({\rm i} q \sigma){\rm d} q$, $q$ being the wavenumber of deformation mode and $\delta h\ll h$ a constant part of cell-height deformation. Upon these transformations, the system of governing equations simplifies to a single polynomial equation, which is solved to obtain the force $\mu$ as a function of the wavenumber $q$. At the onset of instability (i.e., at small $\mu$ and $q$), we obtain
\begin{multline}
    \mu = \frac{K}{q}+\frac{K \left(h ^4+2 h^2  \Delta -2\right)q}{8 h^2 }+ \\ +\frac{\left (8B-\Delta ^2+2\right )q^2}{8}+ \frac{K\left(h ^4+5\right) q^3}{16 h ^4}+ \\ +\frac{ \left[8B\left (h ^4/2+h^2  \Delta -1\right )+h ^4-\Delta ^2\right]q^4}{32 h^2}\>.
    \label{eq:critical_mu}
\end{multline}
Since the total energy equals the work of the applied load $P=\int_0^{Nh^{-1}} \mu {\rm d}\sigma$, the dominant deformation mode at the point of instability is characterized by $q_0$, where $\left (\partial \mu/\partial q\right )_{q=q_0} = 0$.

Throughout the text, the analytical results are compared with the results of vertex-model simulations that minimize the discrete model (Eq.~\ref{eq:sc_energy}) in its full non-linear form~({\color{blue}Materials and Methods}). In all figures, the results of the vertex model are denoted by datapoints, whereas the analytical results are represented by solid curves. 

{\bf Buckling.} Depending on $\Gamma$, $\Delta$, $B$, and $K$, tissues undergo either buckling or wrinkling instability, where $q_0=0$ ($q_0=2\pi/L$ for tissues with finite end-to-end length $L$) or $q_0>0$, respectively. In absence of the apico-basal differential tension~(i.e., $\Delta=0$) and supporting structures (i.e., $B=K=0$), a tissue with an end-to-end length $L$ {\it buckles} at a critical force $\mu_0=\pi ^2/L^2\sim h^{0}$~(SI Appendix). This result agrees with numerical results~(Fig.~\ref{fig:F1}\textit{D}) and is in contrast with Euler instability in solid plates, where $\mu_c\sim h^3$.

\begin{figure}[htb!]
    \centering
    \includegraphics{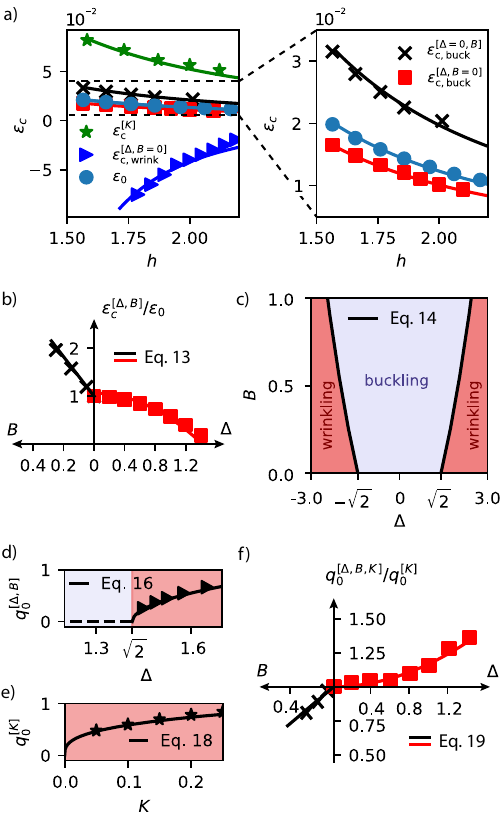}
    \caption{Instability analysis in epithelial sheets. a)~Critical compressive strain $\epsilon_c$ of the epithelium in different limits: Unsupported epithelium with non-polar surface tensions ($\epsilon_0$), epithelium, supported by BM ($\epsilon^{[B]}$), epithelium, supported by stroma ($\epsilon^{[K]}$), and epithelium with differential surface tension ($\epsilon^{[\Delta]}_{\rm buck}$ and $\epsilon^{[\Delta]}_{\rm wrink}$ for buckling and wrinkling deformation mode, respectively). b)~Relative critical compressive strain $\epsilon$ for buckling instability as a function of $B$~(black) and $\Delta$~(red); $\Gamma=3.5$. c)~Phase diagram $B(\Delta)$ showing regions of buckling and wrinkling. d)~Wavenumber of wrinkling in an unsupported epithelium ($B=K=0$) as a function $\Delta$ for $\Gamma=4$. e)~Wavenumber of wrinkling in an epithelium supported by the stroma as a function of $K$ for $\Delta=B=0$ and $\Gamma=4$. f)~Relative optimal wavenumber of wrinkling patterns, $q_0^{[K, \Delta, B]}$, as a function of $B$ (black) and $\Delta$~(red); $\Gamma=3.5$ and $K=0.1$. In all panels, datapoints represent vertex-model results.}
    \label{fig:F2}
\end{figure}

The critical buckling force is related to the critical compressive strain by~(SI Appendix)
\begin{equation}
    \mu_c=2h^2\epsilon_c\>,
\end{equation}
such that in absence of differential tension and supporting structures, 
\begin{equation}
    \epsilon_0=\frac{\pi ^2}{2 h^2L^2}\sim h^{-2}\>.
    \label{eq:crit_strain}
\end{equation}
Interestingly, while $\epsilon_0$, like $\mu_0$, is inversely proportional to the length squared (i.e., $\epsilon_0\sim L^{-2}$), similarly to the standard Euler instability in solid plates, the $h^{-2}$-dependence on the layer thickness in Eq.~\ref{eq:crit_strain} is, again, fundamentally different from the $h^2$-dependence in plates (Eq.~\ref{eq:crit_strain} and Fig.~\ref{fig:F2}\textit{A}). 

Buckling persists when the tissue is supported by BM, however, the critical compressive strain increases relatively to the unsupported case and reads $\epsilon_0\left(1+4B\right)$. In contrast, the apico-basal differential tension $\Delta$ tends to decrease the critical compressive strain as $\epsilon_0 (1-\Delta^2/2)$. At $\Delta=\sqrt{2}$, where the tissue transitions from buckling to wrinkling, as discussed in the next section, the critical compressive strain is 0.

Tissues with non-zero differential tension which are additionally supported by a BM (i.e., $\Delta\neq 0$ and $B>0$), buckle once the compressive strain reaches a critical value~(Fig.~\ref{fig:F2}\textit{B})
\begin{equation}
    \epsilon^{[\Delta, B]}_{c,{\rm buck}}= \epsilon_0\left(1-\frac{\Delta^2}{2}+4B\right)\>,
\end{equation}
which includes both individual contributions of $\Delta$ and $B$.

{\bf Buckling-to-wrinkling transition.} The competing effects of the combined tissue-BM bending rigidity and the apico-basal differential tension result in a transition from buckling to wrinkling, whereby an initially flat tissue becomes contractile and it spontaneously loses stability at a periodic mode, characterized by a wavenumber $q_0$, which minimizes $\mu$~(Eq.~\ref{eq:critical_mu}). The contractile behavior is seen in a negative critical compressive strain for $\left |\Delta\right |$ above the wrinkling threshold~(Fig.~\ref{fig:F2}\textit{C})
\begin{equation}
    \Delta_{c}^{[B]}= \sqrt{2(1+4B)}\>,
    \label{eq:optimal_vector_delta_b}
\end{equation}
where
\begin{equation}
    \epsilon^{[\Delta, B]}_{c,{\rm wrink}}=-\frac{4 B +1}{2\left(h^4-2\right)} \left(\left |\Delta\right | -\Delta_c^{[B]}\right)^2\>.
\end{equation}
In absence of BM (i.e., $B=0$), the critical differential tension reduces to $\Delta_c=\sqrt{2}$, which agrees with the result previously derived for unsupported tissues \cite{andrensek23}. Note that here, both $\epsilon_{c,{\rm wrink}}^{[\Delta,B]}$ and $\mu_{c,{\rm wrink}}^{[\Delta,B]}=2h^2\epsilon_{c,{\rm wrink}}^{[\Delta,B]}$ scale differently with the thickness $h$ compared to the buckling regime~(Fig.~\ref{fig:F2}\textit{A}).

{\bf Wrinkling.} For $\left |\Delta\right |>\Delta_c^{[B]}$, tissues develop periodic wrinkling patterns, characterized by a wavenumber~(Fig.~\ref{fig:F2}\textit{D})
\begin{equation}
    q_0^{[\Delta, B]} = \frac{2^{5/4}}{\sqrt{h^2 -2/h^2}}\left(\left |\Delta\right| -\Delta_c^{[B]}\right)^{1/2}.
    \label{eq:optimal_wave_vector_db}
\end{equation}

The presence of the stroma causes wrinkling regardless of the presence of BM and the differential surface tension, i.e., the optimal wavenumber $q_0$ is always non-zero and the divergence in denominator is only apparent since $\sqrt{2}<\Delta<h^2$. When considering only the effects of the stroma and neglecting both the differential surface tension and the interaction with BM (i.e., $\Delta=B=0$), the corresponding governing equations yield a critical wrinkling compressive strain
\begin{equation}
    \epsilon_c^{[K]}= 3\cdot 2^{-7/3}h^{-2}K^{2/3}\sim h^{-2}\>
\end{equation}
and a critical force $\mu_c^{[K]}= 3\cdot 2^{-4/3}K^{2/3}\sim h^{0}$ with a dominant wrinkling mode, given by a wavenumber~(Fig.~\ref{fig:F2}\textit{E})
\begin{equation}
    q_0^{[K]}=\left(2 K \right)^{1/3}\sim h^{0}\>.
\end{equation}
Scaling relations of $\epsilon_c$, $\mu_c$, and $q_0$ with $K$ are the same as in supported solid plates, but the scaling of all three quantities with thickness $h$ is fundamentally different. Indeed, in solid plates, $\epsilon_c\sim h^0$, $\mu_c\sim h$, and $q_0\sim h^{-1}$ \cite{brau13}.
\begin{figure*}[htb!]
    \centering
    \includegraphics[width=0.95\linewidth]{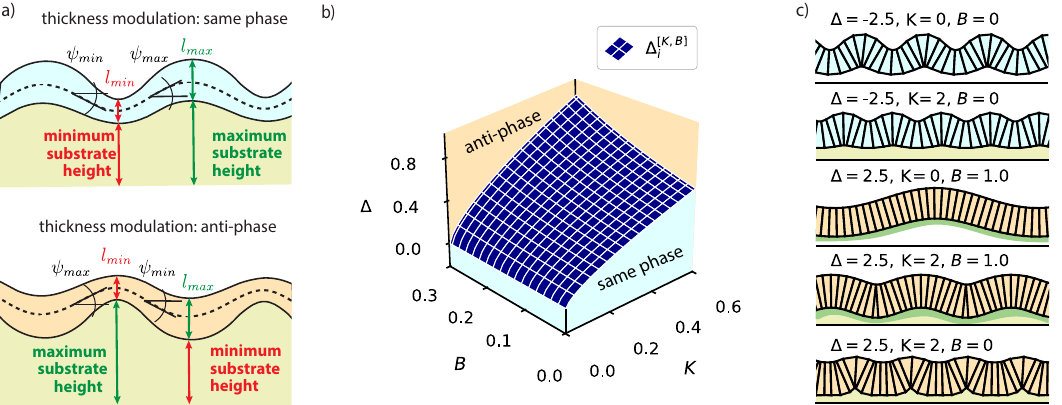}
    \caption{Thickness modulation phase. a)~Schematics of a tissue exhibiting thickness modulation in phase with substrate undulations~(top) and a tissue exhibiting tissue modulation with opposite phase compared to substrate undulations~(bottom). b)~Phase diagram $\Delta(B,K)$ showing regions of in-phase and anti-phase thickness modulations. The dark blue surface shows $\Delta_i^{[\Delta,B,K]}$, where the phase is inverted. c)~Examples of optimal tissue shapes obtained by vertex-model simulations. The cell color denotes whether the thickness modulation has the same phase as substrate undulations~(blue) or an opposite phase~(orange). In all simulations, $\Gamma=4$.}
    \label{fig:F3}
\end{figure*}

In tissues supported by stroma, where, additionally, the effects of $\Delta$ and $B$ are present, we examine the optimal wavenumber perturbatively. Since $K$ in Eq.~\ref{eq:critical_mu} appears at the lowest order in $q$, its influence is greater than that of $\Delta$ and $B$, which is why we can treat corrections due to $\Delta$ and $B$ perturbatively as $q_0^{[\Delta,B,K]}=q_0^{[K]}+\delta q(\Delta,B,K)$, where $\delta q(\Delta,B,K)\ll q_0^{[K]}$. For small $\Delta$ and $B$, we obtain
\begin{equation}
    q_0^{[\Delta, B,K]}=q_0^{[K]}\left(1+ \frac{1}{6}\left[\Delta ^2-8 B \right]\right)\>.
    \label{eq:optimal_wave_vector_k_delta_b}
\end{equation}
This result shows that BM's bending rigidity increases the wavelength while $\left |\Delta\right |$ decreases it~(Fig.~\ref{fig:F2}\textit{F}). The asymmetry between positive and negative $\Delta$-values due to basal presence of the substrate enters $q_0$ at the next lowest order~(SI Appendix). Note that because Eq.~\ref{eq:optimal_wave_vector_k_delta_b} was obtained perturbatively for small deviations of the wavenumber from $q_0^{[K]}$, assuming small $\Delta$- and $B$-values, it does not reduce to Eq.~\ref{eq:optimal_wave_vector_db} for $K=0$.

\section*{Thickness modulation}
The characteristic wavelength of the deformation pattern is often the primary geometric observable in deformed multilayered elastic sheets and can be used to determine the relative stiffnesses of the layers. However, our results show that the elastic interaction among the layers may not always be the dominant mechanism of wrinkling, since wrinkling can also appear in unsupported sheets due to specific internal forces, namely differential interfacial tensions. As a result, the wrinkling wavelength may not be the sole and most reliable observable to pinpoint the dominant mechanism of wrinkling in soft biological tissues.

An additional shape feature of interest is the groove-to-crest modulation of the layer thickness. Indeed, in supported solid plates, the thickness is modulated along a waveform and this modulation is universally {\it in phase} with the undulation of the layer-substrate interface, such that the top (thin) layer is thicker in crests than in grooves \cite{tallinen14, Razavi2015}. In contrast, the modulation phase in epithelia can be inverted and the grooves become thicker than the crests (i.e., the thickness modulates in {\it anti-phase} with substrate undulations)~(Fig.~\ref{fig:F3}\textit{A}). To explain this, Engstrom {\it et al.} \cite{engstrom18} adapted the classical elastic model of supported plates by including the interaction with system-spanning elastic fibers that are present in certain organs. Our model suggests that these large-scale mechanisms are not needed since the phase inversion may naturally emerge from the surface mechanics at the cell level.
\begin{figure*}[htb!]
   \centering
   \includegraphics[width=17.5cm]{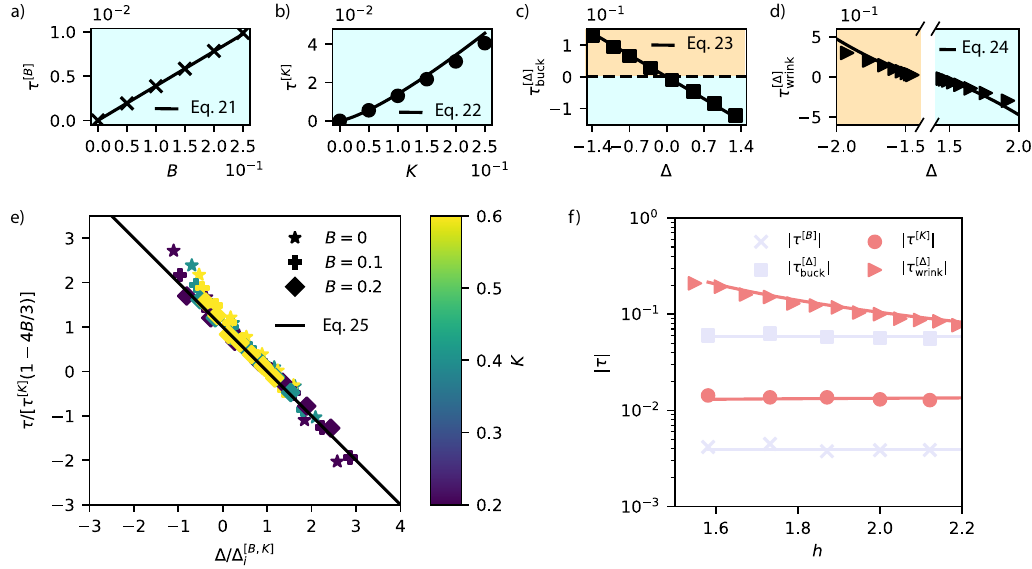}
   \caption{Thickness modulation in epithelial sheets. a),~b)~Thickness modulation $\tau$ as a function of $B$ and $K$ in the limit of no differential tension and only BM and stroma present (a and b, respectively). c),~d)~Thickness modulation $\tau$ as a function of $\Delta$ in the limit of no substrate present in the buckling and the wrinkling regime (c and d, respectively). In a-d, $\Gamma=4$ and blue and orange areas represent same-phase and anti-phase modulation, respectively. e)~Comparison of analytical expression for thickness modulation (solid line) and numerical results of vertex model simulations in tissues with $B=0$, $0<K<0.1$, $-0.1<\Delta<0.3$, and $\Gamma=4$. f)~ Comparison of thickness modulation amplitudes $\left |\tau\right |$ in tissues of similar length with different supporting structures. Parameter values are $B=0.1$ for $\tau^{[B]}$, $K=0.1$ for $\tau^{[K]}$, $\Delta=0.6$ for $\tau^{[\Delta]}_{\rm buck}$, and $\Delta=1.6$ for $\tau^{[\Delta]}_{\rm wrink}$. Lilac and brick colors represent bucklind and wrinkling modes, respectively. In all panels, datapoints represent vertex-model results.}
   \label{fig:F4}
\end{figure*}

In particular, close to the point of elastic instability, governing equations~(SI Appendix) are solved by $l(\sigma)=\delta h + \Lambda\cos(q \sigma)$, $y(\sigma)=Y\cos(q \sigma)$, and $\psi(\sigma)=\Psi\sin(q \sigma)+\Psi_1\cos(q \sigma)$. 
Together with Eq.~[\ref{eq:constraint_boundary}], which connects the amplitudes by $Y=-\Lambda/2 - \Psi/q$, this solution allows a simple definition of thickness modulation, relative to the amplitude of substrate undulations:
\begin{equation}
    \tau=\frac{\Lambda}{Y}\approx -\frac{q \Lambda}{\Psi}\>.
\end{equation}
Its magnitude and the sign (phase) are calculated as $|\tau|$ and ${\rm sgn}(\tau)$, respectively; $\tau$ is defined such that ${\rm sgn}(\tau)=1$ for tissues that are thicker in crests (i.e., thickness modulation and the stroma undulation are in phase), whereas ${\rm sgn}(\tau)=-1$ for tissues thicker in grooves (i.e., thickness modulation and the stroma undulation are in anti-phase)~(Fig.~\ref{fig:F3}\textit{A}).

The ratio of amplitudes $\Lambda/\Psi$ and the wavenumber $q$ follow directly from governing equations and are derived for different regimes in SI Appendix. We find that while an unsupported tissue with no apico-basal polarity (i.e., $K=B=\Delta=0$) exhibits no thickness modulation, i.e., $\tau=0$ , in any other case, the thickness is modulated and this modulation can either be in phase or in anti-phase with substrate undulations, depending on parameters~(Fig.~\ref{fig:F3}\textit{B,~C}).

{\bf In-phase thickness modulation.} In presence of BM but absence of the stroma (i.e., $B\neq 0$, $K=0$), the tissue with no apico-basal differential tension (i.e., $\Delta=0$) bends such that $q^{[B]}=2\pi/L$. In this case, the thickness modulation
\begin{equation}
    \tau^{[B]}=\frac{4 \pi^4 B}{L^4}>0
    \label{eq_tauB}
 \end{equation}
is always in-phase with substrate undulations~(Fig.~\ref{fig:F4}\textit{A}).  

When BM is absent but the epithelium is supported by the stroma (i.e., $B=0$, $K>0$), the tissue wrinkles and the thickness modulation
\begin{equation}
    \tau^{[K]}=2^{-5/3}K^{4/3}>0
    \label{eq_tauK}
\end{equation}
is, again, in phase with substrate undulations~(Fig.~\ref{fig:F4}\textit{B}).

{\bf Phase inversion.} The modulation phase cen be inverted by the apico-basal differential surface tension $\Delta$. In absence of supporting structures (i.e., $B=K=0$), the modulation, caused by $\Delta$, differs between buckling and wrinkling. In a buckled tissue of end-to-end length $L$, it reads
\begin{equation}
    \tau^{[\Delta]}_{\rm buck} = -\frac{\pi^2\Delta }{L^2} \gtrless 0\>,
    \label{eq_tauD}
\end{equation}
whereas in a wrinkled tissue
\begin{align}
    \tau^{[\Delta]}_{\rm wrink}= -\frac{\sqrt{2}\Delta \left(|\Delta| -\sqrt{2}\right)}{h^2-2/h^2}\gtrless 0  \>.
    \label{eq_tauDw}
\end{align}
In both deformation regimes, the value of $\tau$ can be either positive or negative and the phase in both cases changes at $\Delta=\Delta_i^{[\Delta]}=0$~(Fig.~\ref{fig:F4}\textit{C, D}). 

The phase inversion can be intuitively understood by observing the size of apical and basal tissue surfaces in in-phase and in anti-phase configurations (Fig.~\ref{fig:F3}\textit{A}): When thickness modulation is in phase with substrate undulations, the apical area is larger than the basal, whereas the opposite is true in anti-phase configurations. Tissues can, thus, invert the modulation phase by increasing the apical tension, thereby decreasing their apical surfaces.

An inverse of modulation phase can occur even in the presence of supporting structures. While the combined effects of the differential surface tension, BM, and the stroma cause the tissue to wrinkle with a wavelength, predominantly determined by the stiffness of the stroma~(Eq.~\ref{eq:optimal_wave_vector_k_delta_b}), the associated thickness modulation,
\begin{equation}
    \tau^{[K, \Delta, B]}=\tau^{[K]}\left(1 -\frac{4}{3}B-\frac{2^{1/3}\Delta}{K^{2/3}}\left[1-\frac{4}{3}B\right]^2\right)\>,
    \label{eq:modulation_k_delta_b}
\end{equation}
is primarily set by the differential surface tension. In particular, thickness modulation inverts its phase when $\tau$ switches sign, which occurs when the apico-basal differential surface tension exceeds a transition value
\begin{equation}
    \Delta_i^{[B,K]} =  \frac{2^{-1/3}K^{2/3}}{1-4B/3}\>,
\end{equation}
which is entirely determined by elastic properties of the substrate, i.e., BM and the stroma~(Fig.~\ref{fig:F3}\textit{B} and Fig.~\ref{fig:F4}\textit{E}). Note that, similarly to Eq.~\ref{eq:optimal_wave_vector_k_delta_b}, Eq.~\ref{eq:modulation_k_delta_b} is only valid for small $B$, and $\Delta$ (SI Appendix).

In terms of the contribution of various effects to the thickness-modulation amplitude, $\left |\tau\right |$, Fig.~\ref{fig:F4}\textit{F} confirms that the effect of the differential tension is much higher than the effects of the interaction of tissue with either BM or the stroma. Additionally, the thickness-modulation amplitude depends on the layer's baseline thickness $h$ only when the modulation is caused by the apico-basal differential tension, sufficient to drive tissue wrinkling~(Fig.~\ref{fig:F4}\textit{F} and Eqs.~\ref{eq_tauB}-\ref{eq_tauDw}). 
\begin{table*}[htb!]
\centering
\caption{Comparison of our theory predictions with standard elastic (bi)layers.\label{tab1}}
\begin{tabular}{l|l|l|l}
 & {\bf supported epithelia}  & {\bf solid plates} & {\bf supported solid plates} \\
\hline \hline
{\bf instability mode} & buckling/wrinkling & buckling & wrinkling \\
\hline
{\bf wavenumber} & $q=2\pi/L$ for $K=0$ \& $\Delta<\Delta_c^{[B]}$ & $q=2\pi/L$ & $q\sim K^{1/3}h^{-1}$ \\
 & $q\sim\left (\Delta-\Delta_c^{[B]}\right )^{1/2}$ for $K=0$ \& $\Delta>\Delta_c^{[B]}$ &  &  \\
 & $q\sim K^{1/3}h^0$ for $K>0$ &  &  \\
\hline
{\bf critical force} & $\mu_c\sim h^{0}L^{-2}$ for $K=0$ \& $\Delta<\Delta_c^{[B]}$ & $\mu_c\sim h^3L^{-2}$ & $\mu_c\sim K^{2/3}h$ \\
 & $\mu_c\sim\left (\Delta-\Delta_c^{[B]}\right )^2$ for $K=0$ \& $\Delta>\Delta_c^{[B]}$ &  &  \\
  & $\mu_c\sim K^{2/3}h^{0}$ for $K>0$ &  &  \\
\hline
{\bf critical compressive strain} & $\epsilon_c\sim h^{-2}L^{-2}$ for $K=0$ \& $\Delta<\Delta_c^{[B]}$ & $\epsilon_c\sim h^2L^{-2}$ & $\epsilon_c\sim K^{2/3}h^0$ \\
 & $\epsilon_c\sim\left (\Delta-\Delta_c^{[B]}\right )^2$ for $K=0$ \& $\Delta>\Delta_c^{[B]}$ &  &  \\
 & $\epsilon_c\sim K^{2/3}h^{-2}$ for $K>0$ &  &  \\
\hline
{\bf groove-crest} & in phase/in anti-phase & N/A  & in phase  \\
{\bf thickness-modulation phase} & with substrate undulations &  & with substrate undulations \\
\hline
{\bf groove-crest} & $y_0\times f(\Delta,B,K)$ (Eq.~[\ref{eq:modulation_k_delta_b}]) & N/A  & $\ll y_0$ \\
{\bf thickness-modulation amplitude} &  &  &  \\
\hline
\end{tabular}

\end{table*}

\section*{Discussion}
We developed an elasticity theory for epithelial tissues supported by the basement membrane and the stroma. Unlike most existing theories, which assign specific elastic properties to the epithelium \cite{hanezzo11, shyer13vili}, our approach derives these properties from a microscopic, cell-scale model based on differential interfacial tensions. By adopting a simplified mechanical framework that treats cells as incompressible droplets with stresses localized at their surfaces, we examined the influence of interfacial surface tensions on epithelial elasticity. 

We compared  the results with the classic elasticity of solid plates, where stresses are distributed throughout the volume, and identified several key differences, summarized in Table~\ref{tab1}. Firstly, while classic thin plates wrinkle only when supported by a thick substrate, epithelia may wrinkle due to finite-thickness effects and the associated apico-basal tension asymmetry even in absence of substrate. Secondly, the different origin of elasticity results in specific scaling relations of the wavenumber as well as the critical buckling/wrinkling force and compressive strain with layer thickness. In our model, thickness is varied by varying cells' aspect ratio while maintaining fixed cell cross-section areas, however, the same scaling relations are obtained in a scenario where the height is varied by varying the area while fixing the aspect ratio (SI Appendix). Finally, deformed epithelial monolayers exhibit modulation of thickness that is either in phase or in anti-phase with undulations of the epithelium-substrate interface. While the in-phase modulation is not surprising \cite{tallinen14}, the anti-phase modulation is, since it is not present in classic supported solid plates and has been only recently observed in epithelial monolayers \cite{engstrom18}. Importantly, our model predicts that the anti-phase modulation may only arise when the apical tension is greater than the basal tension.

The predicted anomalous elastic properties and deformations of epithelia highlight the need for detailed experimental validation. Our work offers a comprehensive set of testable predictions that can facilitate the characterization of surface mechanics' contributions to the elasticity and morphology of epithelial monolayers, relying on observables that are readily measurable from static images of epithelial cross-sections and can be used to estimate $K$, $B$, $\Gamma\approx h^2$, and $\Delta$. In unsupported buckled tissues, the measurements of thickness modulation $\tau$ and tissue length $L$ can be used to estimate the ratio $B/\Delta$, while an additional measurement of wave-number $q$ can serve to estimate both $B$ and $\Delta$ when an unsupported tissue is wrinkled. In supported wrinkled tissue, the measurements of $q$ and $\tau$ can be used to estimate the ratios $K/\Delta$ and $B/\Delta$, while an additional \textit{in vitro} measurement of critical compressive strain $\epsilon_c$ would be required to estimate $K, B, \Delta,$ and $\Gamma$ independently.

In this study, we neglected cells' bulk elasticity so as to study the effects of surface tensions and to emphasize their contribution to the emergent elasticity as clearly as possible. Experimentally, the relevance of surface tensions in epithelial mechanics may be underscored through a simple dimensional analysis where cell surface energy is compared with the bulk elastic energy, $\gamma L^2\sim EL^3$. This comparison yields the elastocapillary length scale $\xi\sim\gamma/E$, representing a linear cell dimension below which surface tensions dominate over bulk elastic forces. Depending on cell type and various measurements, $\xi$ can be estimated to range between 
$0.1\mu{\rm m}$ and $10\mu{\rm m}$ \cite{wei08, Nussenzveig2018, Fischer-Friedrich2014}. Given that linear dimensions of cells are typically around $1-10\>\mu{\rm m}$, surface forces are not negligible, however they also may not be dominant over bulk elastic forces. Therefore, to better understand the interplay between these two types of forces, future work will need to generalize the current theory to incorporate cells' bulk elasticity as well as to extend it to curved tissues and, lastly, more rigorously include three-dimensional effects \cite{Bal2025}.

Our current theory could readily be tested in systems where entirely neglecting bulk elasticity is more directly justifiable. Some examples include sheets of adherent lipid vesicles or non-biological sheets composed of immiscible microfluidic droplets \cite{kintseS2010, Kamiya2016}. In these setups, differential surface tensions could be controlled by anchoring the sheets at interfaces between immiscible liquids or by introducing surfactants to modulate interfacial properties. Together with the insights from our theory, these proposed experiments may prove useful within a broader field of the study of soft metamaterials, which seeks to understand how design properties at smaller scales influence emergent elastic properties at the system level. Within this context, our theory highlights that microscopic interactions can have a profound impact on the overall properties, deformations, and shapes of elastic sheets, which may exhibit behaviors distinct from conventional elastic layers due to the unique microscopic origins of their elasticity.

\section*{Materials and methods}
\subsection{Vertex model}
Our vertex model describes the tissue as a chain of quadrilateral cell cross sections. The basal, apical, and lateral edges of cell $i$ are under tensions $\Gamma_b$, $\Gamma_a$, and $\Gamma_l$, respectively, and their lengths are denoted by $\ell_i^b$, $\ell_i^a$, and $l_i$, respectively. Additionally, the tissue is supported by a basement membrane~(BM) with a bending rigidity $B$ and the stroma with a bulk elastic modulus $\bar K$. The dimensionless potential energy of the epithelium-BM-stroma system reads
\begin{align}
    \nonumber e=\sum_i^{N}\Big [\kappa_A\left (A_i-1\right )^2+\frac{\Gamma+\Delta}{2} \ell_i^a+\frac{\Gamma-\Delta}{2} \ell_i^b+\\+l_i  +\frac{B}{2}\rho_i^{-2}\frac{\ell_{i-1}^b+\ell_i^b}{2}+\frac{\bar K}{2}y_{i,b}^2\ell_i^b\Big ]\>,
\end{align}
where $\rho_i$ and $y_{i,b}$ denote the local curvature radius of the basement membrane and the $y$-coordinate of the basal vertex of $i$-th lateral edge, respectively. The cell-area modulus $\kappa_A=100$, describes nearly incompressible cells with actual areas close to the preferred $A=1$. All edge lengths and cell areas are calculated as $l_i=\left |\boldsymbol r_{i,{\rm head}}-\boldsymbol r_{i,{\rm tail}}\right |$ and $A_i=(1/2)\sum_\mu\left (\boldsymbol r_{i,\mu}\times\boldsymbol r_{i,\mu+1}\right )\cdot\left (0,0,1\right )$, respectively, where $\mu$ denotes a internal index of the cells, $\mu=1...4$, whereas the local squared curvature of the basement membrane is calculated as $\rho_i^{-2}=8 (1+\cos{\theta_i})/|r_{i+1, {\rm basal}} - r_{i-1, {\rm basal}}|^2$; the expression for angle $\theta_i$ is given in SI Appendix. The energy is minimized using gradient descent, such that each vertex $\boldsymbol r_j=(x_j,y_j)$ moves with a velocity, proportional to the sum of conservative forces given by $\boldsymbol F_j=-\nabla_je$.

In each simulation, we treat only one period of a periodic pattern, assuming periodic boundary conditions in the horizontal direction, such that $N$ in $e$ denotes the number of cells per waveform. Therefore, $q\approx 2\pi/L_0$, where $L_0=Nh^{-1}$ is the initial length of the simulation box and, thus, $\bar K=2\pi hK/N$. To find the optimal number of cells per waveform and the critical compressive strain, at which the elastic instability occurs, our simulations both vary the cell number $N$ and the compressive strain $\epsilon=\delta L/L_0$ as well as simulate the descent of the system towards the minimium of $e$ so as to find the exact optimal tissue shape.

\section{Software availability}
The software used to compute all numerical data included in the manuscript is available at \url{https://github.com/uandrensek/2d-cross-section-vertex-model}.
\bibliographystyle{naturemag}
\bibliography{apssamp}

\vspace{1cm}
\textbf{Author contributions. } MK designed the research, UA performed the analytical calculations and numerical simulations. UA and MK wrote the manuscript.

\textbf{Author declaration. }No competing interests.

\begin{acknowledgments}
We thank Marko Popovic, Michael Moshe, and all the members of the Theoretical biophysics group at JSI for fruitful discussions. We thank Primo\v z Ziherl and Tanmoy Sarkar for critically reading the manuscript. We acknowledge the financial support from the Slovenian Research and Innovation Agency~(research projects J1-3009 and J1-60013, development funding pillar RSF-0106, and research core funding No. P1-0055). This research was also supported in part by grant no. NSF PHY-2309135 to the Kavli Institute for Theoretical Physics (KITP).
\end{acknowledgments}

\end{document}



\title{Emergent epithelial elasticity governed by\\interfacial surface mechanics and substrate interaction:\\Supplemental material}

\author{Ur\v ska Andren\v sek}
\email{urska.andrensek@ijs.si}
\affiliation{Faculty of Mathematics and Physics, University of Ljubljana, Jadranska 19, SI-1000 Ljubljana, Slovenia}
\affiliation{Jo\v zef Stefan Institute, Jamova 39, SI-1000 Ljubljana, Slovenia}%

\author{Matej Krajnc}
\affiliation{Jo\v zef Stefan Institute, Jamova 39, SI-1000 Ljubljana, Slovenia}%

\maketitle
\onecolumngrid

\section{From discrete model to continuum theory}
\subsection{Geometry}
\begin{figure}[htb!]
    \centering
    \includegraphics[]{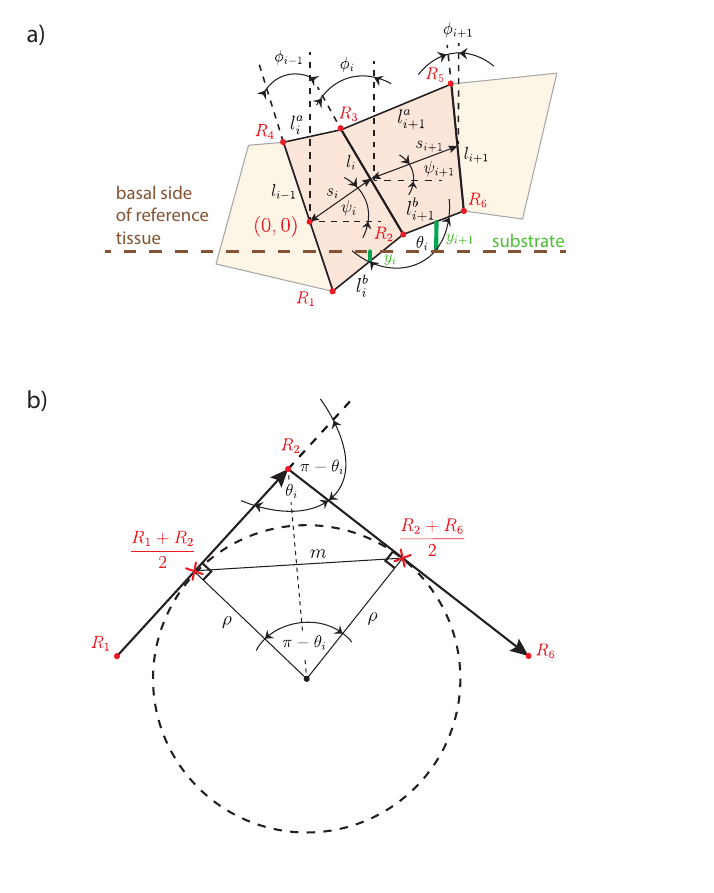}
    \caption{a) Geometry of two adjacent cells. b) Local geometry at vertex ${\bf R_2}$ and the local radius of curvature $\rho_i$.}
    \label{SI:fig_1}
\end{figure}

We begin the derivation of the continuum theory at the cell level by parametrizing the geometry of a pair of neighboring cells~(Fig. \ref{SI:fig_1}). The cells are quadrilateral and they have equal and constant cross-section areas $A_0$. The shape of each cell is parametrized by in-plane positions of its four vertices. The coordinates of the vertices are expressed with discrete variables $\phi$, $\psi$, $l$, and $s$ as
\begin{equation}
    \begin{aligned}
    \boldsymbol R_1&=\frac{1}{2} l_{i-1} \left(\sin (\phi_{i-1}), -\cos (\phi_{i-1}),0\right),\\
    \boldsymbol R_2&=\frac{1}{2} l_i \left(\sin (\phi_i),-\cos (\phi_i),0\right)+\boldsymbol v_1, \\ 
    \boldsymbol R_3&=\frac{1}{2} l_i\left(-\sin (\phi_i),\cos (\phi_i),0\right)+\boldsymbol v_1,\\
    \boldsymbol R_4&=\frac{1}{2} l_{i-1} \left(-\sin (\phi_{i-1}),\cos (\phi_{i-1}),0\right),\\         
    \boldsymbol R_5&=\frac{1}{2} l_{i+1} \left(-\sin (\phi_{i+1}),\cos (\phi_{i+1}),0\right)+\boldsymbol v_2,\\
    \boldsymbol R_6&=\frac{1}{2} l_{i+1}\left(\sin (\phi_{i+1}),-\cos (\phi_{i+1}),0\right)+\boldsymbol v_2\>,
    \label{SI:cell_vertices}
    \end{aligned}
\end{equation}
where
\begin{equation}
    \boldsymbol v_1=s_i \left(\cos (\psi_i),\sin (\psi_i),0\right)
\end{equation}
and
\begin{equation}
    \boldsymbol v_2=s_{i+1} \left(\cos (\psi_{i+1}),\sin (\psi_{i+1}),0\right)+\boldsymbol v_1\>.
\end{equation}

\subsection{Constant-area constraint}
The units of energy and length are given by $\Gamma_l\sqrt{A_0}$ and $\sqrt{A_0}$, such that cell cross-section areas are all equal to 1. We express variables $s_i$ and $s_{i+1}$ from the condition of fixed cell cross-section area. For cells $i$ and $i+1$ this condition reads
\begin{equation}
    A_i=\left (\boldsymbol R_1\times\boldsymbol R_2 + \boldsymbol R_2\times\boldsymbol R_3 + \boldsymbol R_3\times\boldsymbol R_4 + \boldsymbol R_4\times\boldsymbol R_1\right )\cdot\boldsymbol e_z=1\>
\end{equation}
and
\begin{equation}
    A_{i+1}=\left (\boldsymbol R_2\times\boldsymbol R_6 + \boldsymbol R_6\times\boldsymbol R_5 + \boldsymbol R_5\times\boldsymbol R_3 + \boldsymbol R_3\times\boldsymbol R_2\right )\cdot\boldsymbol e_z=1\>,
\end{equation}
respectively. Thus, the midline lengths  
\begin{equation}
    s_i=\frac{2}{l_{i-1} \cos (\phi_{i-1}-\psi_i)+l_i \cos (\phi_i-\psi_i)}
\end{equation}
and
\begin{equation}
    s_{i+1}=\frac{2}{l_{i} \cos (\phi_{i}-\psi_{i+1})+l_{i+1} \cos (\phi_{i+1}-\psi_{i+1})}\>.
\end{equation}

\subsection{Continuum limit}
\label{SI:sec_continuum}
For the reference state, we take a flat configuration, where the stroma and the basement membrane are both undeformed. The surface energy is minimized when the cell height 
\begin{equation}
    \label{SI:hh}
    h=\Gamma^{1/2}
\end{equation}
and the width $h^{-1}=\Gamma^{-1/2}$.

We perform the continuum limit of a given expression $E_i=E_i\left (\psi_{i}, \psi_{i+1}, \phi_{i-1}, \phi_i, \phi_{i+1}, \delta l_{i-1}, \delta l_i, \delta l_{i+1}, y_i\right )$ using the following procedure:
\begin{enumerate}
    \item The expression $E_i$ is expressed as a function of variables, whose value is zero in the reference state. These variables are: $\psi_{i}$, $\psi_{i+1}$, $\phi_{i-1}$, $\phi_i$, $\phi_{i+1}$, $\delta l_{i-1}=l_{i-1}-h$, $\delta l_i=l_{i}-h$, $\delta l_{i+1}=l_{i+1}-h$, $y_i$, and $y_{i+1}$.
    \item The expression $E_i$ is truncated to second order in all the variables.
    \item The discrete variables are replaced by continuous functions in the frame of the reference state, parametrized by $\sigma$:
    \begin{equation}
    \begin{aligned}
        \psi_i &\rightarrow \psi(\sigma) \\
        \psi_{i+1} &\rightarrow \psi(\sigma)+ \dot{\psi}(\sigma)h^{-1} +\frac{1}{2}\ddot{\psi}(\sigma)h^{-2} \\
        \phi_i &\rightarrow \phi(\sigma) \\
        \phi_{i+1} &\rightarrow \phi(\sigma) + \dot{\phi}(\sigma)h^{-1} + \frac{1}{2}\ddot{\phi}(\sigma)h^{-2}\\
        \phi_{i-1} &\rightarrow \phi(\sigma) - \dot{\phi}(\sigma)h^{-1} +\frac{1}{2}\ddot{\phi}(\sigma)h^{-2} \\
        \delta l_i=l_i -h &\rightarrow \delta l(\sigma) \\
        \delta l_{i+1}=l_{i+1} -h &\rightarrow \delta l(\sigma) + \delta \dot{l}(\sigma)h^{-1} + \frac{1}{2}\delta \ddot{l}(\sigma)h^{-2} \\
        \delta l_{i-1}=l_{i-1} -h&\rightarrow  \delta l(\sigma) - \delta \dot{l}(\sigma)h^{-1} + \frac{1}{2}\delta \ddot{l}(\sigma)h^{-2}, \\
        y_i &\rightarrow y(\sigma), \\
        y_{i+1} &\rightarrow y(\sigma) +\dot{y}(\sigma)h^{-1} +\frac{1}{2}\ddot{y}(\sigma)h^{-2}\;.
    \end{aligned}
    \label{SI:substitution}
    \end{equation}
    Here the "dot" denotes the derivative with respect to $\sigma$. To simplify the equations, we hereafter omit "$\delta$" from $\delta l(\sigma)$.
    \item To replace the discrete summation of $E_i$ over all cell indices $i$ with the integral of continuous $W(\sigma)$ along reference-frame coordinate $\sigma$, $E_i$ is multiplied by $h{\rm d}\sigma$ so that
    \begin{equation}
        \sum_{i=1}^N E_i \rightarrow\int_0^{N\sigma_0}he(\sigma){\rm d}\sigma \;,
    \end{equation}
    where $e(\sigma)$ represents the line density of expression $E$.
\end{enumerate}

\subsection{Boundary condition}
We are interested in globally flat tissue configuration with a fixed end-to-end distance $Nh^{-1}\left (1-\epsilon\right )$, where $\epsilon$ is the compressive strain imposed by an externally applied force. This boundary condition implies that the projection of the tissue midline on the horizontal $x-$axis
\begin{equation}
    \sum_{i=1}^N \delta X_i = N h^{-1} (1- \epsilon)\;,
\end{equation}
where
\begin{equation}
    \delta X_i= \frac{1}{2}\left[s_i \cos(\psi_i)+s_{i+1}\cos(\psi_{i+1})\right]\;.
    \label{SI:bc_discreete}
\end{equation}  
Following the steps defined in Sec.~\ref{SI:sec_continuum} leads to an expression
\begin{equation}
    \begin{aligned}
    \delta X_i \rightarrow h\delta x(\sigma)&= \frac{1}{16 \Gamma ^{3}}[16 \Gamma ^3-4 \Gamma ^{3/2} \ddot l+8 \Gamma  l \left(\ddot l-2 \Gamma ^{3/2}\right)+\ddot l^2+4 \Gamma  \dot l^2+16 \Gamma ^2 l^2-2 \Gamma ^{3/2} \dot\psi  \ddot\phi +\\
    &+8 \Gamma ^3 \phi ^2-4 \Gamma ^2 \phi  \left(2 \sqrt{\Gamma } \dot\psi +4 \Gamma  \psi +\ddot\psi -\ddot\phi \right)-4 \Gamma ^2 \psi  \ddot\phi-4 \Gamma ^2 \dot\psi  \dot \phi +\\
    &+4 \Gamma ^2 \dot\phi ^2+\Gamma  \ddot\phi ^2-\Gamma  \ddot\psi  \left(2 \sqrt{\Gamma } \dot\phi +\ddot\phi \right)]\>.
    \end{aligned}
\end{equation}
We recognize the variable $h \delta x$ to be $\dot x$, where $x$ is the projection of tissue midline to horizontal axis. The boundary condition can then be expressed as
\begin{equation}
    Nh^{-1}(1-\epsilon)=\int_0^{N\sigma_0} \dot x {\rm d}\sigma\>.
\end{equation}
and, finally, the contribution to the Lagrangian density, that will assure this condition is satisfied, reads
\begin{equation}
    \mathcal{L}_{C1}=\mu\left(1-\epsilon-\dot x\right)\>,
\end{equation}
where $\mu$ is a Lagrangian multiplier.

\subsection{Surface tension}
The surface-energy contribution from the two cells reads
\begin{equation}
    W_i^s=\frac{1}{4} (l_{i-1}+2 l_i+l_{i+1})+\frac{1}{4} (\Gamma +\Delta ) (\ell_i^a+\ell_{i+1}^a)+\frac{1}{4} (\Gamma -\Delta ) (\ell_i^b+\ell_{i+1}^b)\;,
\end{equation}
where the apical and basal edge lengths, $\ell_i^a$, $\ell_{i+1}^a$, $\ell_i^b$, and $\ell_{i+1}^b$ are calculated from the coordinates of cell vertices~(Eq.~[\ref{SI:cell_vertices}]) as
\begin{equation}
    \begin{aligned}
    \ell_i^a&=|\boldsymbol{R}_4 - \boldsymbol{R}_3|\;,\\
    \ell_{i+1}^a&=|\boldsymbol{R}_3 - \boldsymbol{R}_5| \;,\\
    \ell_i^b&=|\boldsymbol{R}_1 - \boldsymbol{R}_2|\;,\\
    \ell_{i+1}^b&=|\boldsymbol{R}_2 - \boldsymbol{R}_6|\;.\\
    \end{aligned}
\end{equation}
Performing the continuum limit as described in Sec.~\ref{SI:sec_continuum} and eliminating terms that give 0 upon integration results in
\begin{equation}
    \begin{aligned}
    W_i^s \rightarrow w_s &=\frac{\Delta  \ddot l \ddot  \psi }{8 \Gamma ^{3/2}}+\frac{\ddot l^2}{8 \Gamma ^2}+ \frac{(4 l \ddot l+\ddot \psi )^2-\ddot \psi  \ddot \phi +\ddot \phi ^2}{8 \Gamma } +\frac{2 \Delta  \dot l \dot \psi +2\dot \psi  \ddot \psi}{4 \sqrt{\Gamma }}+\Gamma  \left(\frac{1}{4} \dot l^2+\psi ^2-2 \psi  \phi+\phi ^2+4\right)+\\ &+\frac{1}{16} \left(\ddot l^2+16 \Delta  \psi \dot l+32 l^2-8 \psi  \ddot \phi -8 \dot \psi  \dot \phi -8 \phi  \ddot \psi \right)+\sqrt{\Gamma } \left(\dot \psi  (\psi -\phi) -\Delta  \dot \phi\right)\;.
    \end{aligned}
\end{equation}
The surface-energy contribution to the Lagrangian density is then
\begin{equation}
    \mathcal{L}_S=w_s\;.
\end{equation}

\subsection{Basement membrane}
The bending-energy density of the basement membrane with bending modulus $B$ depends on the local curvature of the membrane. The associated bending energy of a line segment around vertex ${\bf R_2}$ reads
\begin{equation}
    W_i^B=\frac{B}{2}\left(\frac{1}{\rho_i}\right)^2\, \frac{\ell_i^b+\ell^b_{i+1}}{2}\;,
\end{equation}
where $\rho_i$ is the local radius of curvature at vertex $\boldsymbol{R}_2$, determined by fitting a circle with a diameter $2\rho_i$ to midpoints of edges with length $\ell_i^b$ and $\ell_{i+1}^b$ (Fig. \ref{SI:fig_1}b). The corresponding local curvature $1/\rho_i$ is then expressed as 
\begin{equation}
    \frac{1}{\rho_i}=\frac{2}{m}\sin\left(\frac{\pi-\theta_i}{2}\right)\;,
\end{equation}
where $m$ is the distance between the midpoints of basal edges enclosing the angle $\theta_i$. By using trigonometric relations $\sin(\pi/2 -x) = \cos(x)$ and $\cos^2(x/2)=(1+\cos(x))/2$, we obtain for the squared local curvature
\begin{equation}
    \left(\frac{1}{\rho_i}\right)^2=\frac{2[1+\cos(\theta_i)]}{m^2}\;.
\end{equation}
Lastly, the angle $\theta_i$ and the distance $m$ are calculated from positions of vertices as
\begin{equation}
    \cos(\theta_i)=-\, \frac{(\boldsymbol{R}_2 - \boldsymbol{R}_1) \cdot (\boldsymbol{R}_6 - \boldsymbol{R}_2)}{|\boldsymbol{R}_2 - \boldsymbol{R}_1| |\boldsymbol{R}_6 - \boldsymbol{R}_2|}
\end{equation}
and
\begin{equation}
    m^2=\left|\frac{\boldsymbol{R}_2 + \boldsymbol{R}_1}{2} - \frac{\boldsymbol{R}_6 + \boldsymbol{R}_2}{2}\right|^2 = \left|\frac{\boldsymbol{R}_1+ \boldsymbol{R}_6}{2}\right|^2\;,
\end{equation}
respectively. Following the steps defined in Sec. \ref{SI:sec_continuum} gives
\begin{equation}
    W_B \rightarrow {w}_B = \frac{B \left(\ddot l-2 \dot \psi \right)^2}{8}
\end{equation}
and the contribution to Lagrangian density caused by the basement membrane in 
\begin{equation}
    \mathcal{L}_B=w_B\;.
\end{equation}

\subsection{Stroma}
The elastic energy of the stroma per single cell is calculated as
\begin{equation}
    W_i^K=\frac{\bar K(q)}{2}\frac{y_i^2 + y_{i+1}^2}{2}\;,
\end{equation}
where $y_i$ and $y_{i+1}$ are the height differences between the basal side of the reference flat state and the actual positions of the midpoints of basal edges $i$ and $i+1$, respectively (see Fig. \ref{SI:fig_1}a). We use the continuum limit, following the procedure defined in Sec.~\ref{SI:sec_continuum} to obtain
\begin{equation}
    W_i^K \rightarrow w_K = \bar K(q)\,\frac{\left(\ddot y+2 h \dot y+2 h  y\right)^2+4 h^4 y^2}{16 h^{4}}\>.
\end{equation}
The contribution to Lagrangian density is
\begin{equation}
    \mathcal{L}_K=w_K.
\end{equation}

Since the variable $y$ is not independent from other variables, we need to consider an additional constraint that relates $y$ and it's derivatives to variables $\psi$, $l$, and $\phi$ and their derivatives. The deflection of the stroma from its reference undeformed state at cell $i$ can be expressed as
\begin{equation}
    y_i=\sum_{j=1}^is_j \sin(\psi_j) - \frac{l_i}{2}\cos(\psi_i)\;.
    \label{SI:stroma_constraint_discrete}
\end{equation}
This condition can be coarse-grained by following the procedure defined in Sec.~\ref{SI:sec_continuum} with an exception that the discrete expression is here expanded only to the first order -- this simplification will be {\it post-hoc} justified when Euler-Lagrange equations are solved in Sec.~\ref{SI:sec_EL}. The continuum limit yields the following condition
\begin{equation}
    y(\sigma) + \frac{1}{2}\, l(\sigma) - \int_0^{\sigma}\psi(\sigma^*){\rm d}\sigma^* = 0\;,
\end{equation}
which, when differentiated with respect to $\sigma$, yields a constraint
\begin{equation}
    \dot y=\psi - \frac{1}{2}\, \dot l\;.
    \label{SI:stroma_constraint}
\end{equation}
This constraint is imposed by adding the following term to the Lagrangian density
\begin{equation}
    \mathcal{L}_{C2}=Q(\sigma)h\left [\dot y-\psi + \frac{1}{2}\dot l\right ]\;,
    \label{SI:LC2}
\end{equation}
where $Q(\sigma)$ is a Lagrangian multiplier. In Sec. \ref{SI:sec_EL} we confirm that $Q(\sigma)\propto \dot y$, meaning that the term \ref{SI:LC2} is of quadratic order in chosen variables and that expanding expression \ref{SI:stroma_constraint_discrete} only to the first order was justified.

\section{Euler-Lagrange equations}
\label{SI:sec_EL}
The system of Euler-Lagrange equations, derived from the total Lagrangian density
\begin{equation}   \mathcal{L}=\mathcal{L}_S+\mathcal{L}_B+\mathcal{L}_K+\mathcal{L}_{C1}+\mathcal{L}_{C2}\;,
    \label{SI:Lagrangian}
\end{equation}
reads
\begin{align}
    &\begin{aligned}
    &4 \Gamma ^{5/2} \left[\sqrt{\Gamma } \left(\Gamma  \ddot l +2 \Delta  \dot \psi  \right)+4 \mu +\Gamma  \dot Q +\Delta  \ddot \psi  \right]+8 \Gamma ^3  B  \psi ^{(3)} = \\
    &=l^{(4)}  \left(4 \Gamma ^3  B +\Gamma ^3+2 \Gamma +2 \mu \right)+8 \Gamma  (\Gamma +\mu ) \ddot l +32 \Gamma ^2 (\Gamma +\mu ) l +\Gamma ^{3/2} \Delta  \psi ^{(4)} ,  
    \label{SI:EL_equation_1} 
    \end{aligned} \\
    &(\Gamma +\mu ) \left[16 \Gamma ^2 \phi  -4 \Gamma  \left(2 \sqrt{\Gamma } \dot \psi  +4 \Gamma  \psi  +\ddot \psi  \right)-\psi ^{(4)} +2 \phi ^{(4)} \right]=0, 
    \label{SI:EL_equation_2} \\
    &\begin{aligned}
    &4 \Gamma  \left[\sqrt{\Gamma } \Delta \ddot  l +4 \Gamma   B  \ddot \psi  +(\Gamma +\mu ) \ddot \phi  +4 \Gamma  (\Gamma +\mu ) \phi  \right]+8 \Gamma ^{5/2} Q +(\Gamma +\mu ) \phi ^{(4)} =\\
    &=8 \Gamma ^2 \Delta  \dot l +\sqrt{\Gamma } \left[\Delta  l^{(4)} +8 \Gamma ^{3/2}  B  l^{(3)} +8 \Gamma  (\Gamma +\mu ) \dot \phi \right]+16 \Gamma ^3 \psi  +2 \Gamma  \psi ^{(4)} ,
    \label{SI:EL_equation_3} 
    \end{aligned} \\
    &4 \Gamma ^{5/2} \dot Q =K q \left(y^{(4)} +8 \Gamma ^2 y \right)\>.
    \label{SI:EL_equation_4}   
\end{align}
This system can be reduced to a system of three differential equations by differentiating Eq.~\ref{SI:EL_equation_3} with respect to $\sigma$, using Eq.~\ref{SI:EL_equation_4} to eliminate variable $\dot Q$, and once again differentiating Eqs.~\ref{SI:EL_equation_1} and \ref{SI:EL_equation_3} and using Eq.~\ref{SI:stroma_constraint} to eliminate all terms containing $\dot y$. What remains is a system of three differential equations for variables $\psi(\sigma)$, $\phi(\sigma)$, and $l(\sigma)$: 
\begin{align}
        &\begin{aligned}
        &4 \Gamma ^2 \left[\frac{1}{4}\Gamma^{-3/2}  K \left[\left(\psi ^{(4)} -\frac{1}{2} l^{(5)} \right)+8 \Gamma ^2 \left(\psi  -\frac{\dot l }{2}\right)\right]+\sqrt{\Gamma } \left(\Gamma  l^{(3)} +2  B  \psi ^{(4)} \right)+2 \sqrt{\Gamma } \Delta  \ddot \psi  +\Delta  \psi ^{(3)} \right]=\\
        &=l^{(5)}  \left[\sqrt{\Gamma } \left(4 \Gamma ^2  B +\Gamma ^2+2\right)+2 \mu \right]+8 \Gamma  \left(\sqrt{\Gamma }+\mu \right) l^{(3)} +32 \Gamma ^2 \left(\sqrt{\Gamma }+\mu \right) \dot l +\Gamma  \Delta  \psi ^{(5)} ,
        \label{SI:EL_simplified_1}
        \end{aligned} \\ 
        &\left(\sqrt{\Gamma }+\mu \right) \left(16 \Gamma ^2 \phi  -4 \Gamma  \left(2 \sqrt{\Gamma } \dot \psi  +4 \Gamma  \psi  +\ddot \psi  \right)-\psi ^{(4)} +2 \phi ^{(4)} \right)=0, 
        \label{SI:EL_simplified_2} \\ 
        &\begin{aligned}&4 \Gamma  \left[\frac{1}{2} \Gamma^{-3/2}  K \left[\left(\psi ^{(4)} -\frac{1}{2} l^{(5)} \right)+8 \Gamma ^2 \left(\psi  -\frac{\dot l }{2}\right)\right]+\Delta  l^{(4)} +4 \sqrt{\Gamma }  B  \psi ^{(4)} +\left(\sqrt{\Gamma }+\mu \right) \left(4 \Gamma  \ddot \phi  +\phi ^{(4)} \right)\right]+\left(\sqrt{\Gamma }+\mu \right) \phi ^{(6)} =\\
        &=\Delta  l^{(6)} +2 \sqrt{\Gamma } \left(4 \Gamma  \left[ B  l^{(5)} +\left(\sqrt{\Gamma }+\mu \right) \phi ^{(3)} \right]+\psi ^{(6)} \right)+8 \Gamma ^{3/2} \Delta  l^{(3)} +16 \Gamma ^{5/2}\ddot  \psi  .
        \label{SI:EL_simplified_3}
        \end{aligned}
\end{align}
Using the following {\it ansatz}:
\begin{equation}
    \begin{aligned}
    \psi(\sigma)&=\Psi \exp({\rm i} q \sigma), \\   
    \phi(\sigma)&=\Phi \exp({\rm i} q \sigma), \\
    l (\sigma)&=\delta h+\Lambda \exp({\rm i} q \sigma)\>,
    \label{SI:ansatz}
    \end{aligned}
\end{equation}
allows eliminating the amplitudes, defined in Eq.~\ref{SI:ansatz}, and the above system reduces to a single polynomial equation for $q$ that reads
\begin{equation}
    \begin{aligned}
        &3 K \Gamma ^2 q^{15}+6 \Gamma ^4 q^{14}+12 \Gamma ^2 q^{14}-4 \Gamma ^2 \Delta ^2 q^{14}+24 \Gamma ^4  B  q^{14}+16 K \Gamma ^3 q^{13}+16 K \Gamma  q^{13}+40 \Gamma ^5 q^{12}-16 \Gamma ^3 q^{12}-\\
        &-32 \Gamma ^3 \Delta ^2 q^{12}+128 \Gamma ^5  B  q^{12}+128 \Gamma ^3  B  q^{12}+72 K \Gamma ^4 q^{11}-64 K \Gamma ^2 q^{11}+64 K \Gamma ^3 \Delta  q^{11}+\\
        &+96 \Gamma ^6 q^{10}+128 \Gamma ^4 q^{10}-96 \Gamma ^4 \Delta ^2 q^{10}+384 \Gamma ^6  B  q^{10}-512 \Gamma ^4  B  q^{10}+512 \Gamma ^5 \Delta   B  q^{10}+\\
        &+256 K \Gamma ^5 q^9+512 K \Gamma ^3 q^9+256 \Gamma ^7 q^8+512 \Gamma ^5 q^8-512 \Gamma ^5 \Delta ^2 q^8+1024 \Gamma ^7  B  q^8+3072 \Gamma ^5  B  q^8+\\
        &+640 K \Gamma ^6 q^7-1024 K \Gamma ^4 q^7+1024 K \Gamma ^5 \Delta  q^7+512 \Gamma ^8 q^6-512 \Gamma ^6 \Delta ^2 q^6+2048 \Gamma ^8  B  q^6-4096 \Gamma ^6  B  q^6+4096 \Gamma ^7 \Delta   B  q^6+\\
        &+1024 K \Gamma ^7 q^5+5120 K \Gamma ^5 q^5+4096 \Gamma ^7 q^4-2048 \Gamma ^7 \Delta ^2 q^4+16384 \Gamma ^7  B  q^4+2048 K \Gamma ^8 q^3-4096 K \Gamma ^6 q^3+\\
        &+4096 K \Gamma ^7 \Delta  q^3+16384 K \Gamma ^7 q+(-K \Gamma  q^{15}-2 \Gamma ^3 q^{14}+8 \Gamma  q^{14}-8 \Gamma ^3  B  q^{14}+8 K \Gamma ^2 q^{13}+16 K q^{13}+8 \Gamma ^4 q^{12}+\\
        &+32 \Gamma ^2 q^{12}+64 \Gamma ^4  B  q^{12}+128 \Gamma ^2  B  q^{12}-56 K \Gamma ^3 q^{11}-64 K \Gamma  q^{11}-32 \Gamma ^5 q^{10}-256 \Gamma ^3 q^{10}-384 \Gamma ^5  B  q^{10}-\\
        &-512 \Gamma ^3  B  q^{10}+128 K \Gamma ^4 q^9+512 K \Gamma ^2 q^9-256 \Gamma ^6 q^8+2048 \Gamma ^4 q^8+512 \Gamma ^6  B  q^8+3072 \Gamma ^4  B  q^8-640 K \Gamma ^5 q^7-\\
        &-1024 K \Gamma ^3 q^7-5120 \Gamma ^5 q^6-2048 \Gamma ^7  B  q^6-4096 \Gamma ^5  B  q^6+512 K \Gamma ^6 q^5+5120 K \Gamma ^4 q^5-2048 \Gamma ^8 q^4+12288 \Gamma ^6 q^4+\\
        &+16384 \Gamma ^6  B  q^4-2048 K \Gamma ^7 q^3-4096 K \Gamma ^5 q^3-16384 \Gamma ^7 q^2+16384 K \Gamma ^6 q) \mu + \\ 
        &+(-4 q^{14}+48 \Gamma  q^{12}-384 \Gamma ^2 q^{10}+1536 \Gamma ^3 q^8-5120 \Gamma ^4 q^6 +8192 \Gamma ^5 q^4-16384 \Gamma ^6 q^2) \mu ^2=0\>.
    \end{aligned}
\label{SI:polinom}
\end{equation}

\section{Critical force}

To calculate the critical conditions to induce an elastic instability in a uniaxially deformed monolayer, we turn to Eq.~\ref{SI:EL_equation_1}. Assuming the {\it ansatz} specified by Eq.~\ref{SI:ansatz} and noting that the variable $\dot Q$ is harmonic (see Eq.~\ref{SI:EL_equation_4}), we obtain a solution for the relative change of monolayer's thickness
\begin{equation}
    \delta h = \frac{\sqrt{\Gamma } \mu }{2 \Gamma +2 \mu } \approx \frac{\mu}{2 \sqrt{\Gamma}}\>.
    \label{SI:height_force}
\end{equation}

In a flat configuration, the relative change of the cell height is trivially related to the compressive strain $\epsilon$:
\begin{equation}
    \delta h = h \epsilon\>.
    \label{SI:height_change}
\end{equation}
At the critical degree of compression, the tissue transitions from flat to a buckled or a winkled configuration. Combining Eqs.~\ref{SI:height_change} and \ref{SI:height_force} gives a relation between the cirtical force and the critical compressive strain. This relation reads
\begin{equation}
    \mu_c= 2 h^2 \epsilon_c\>.
\end{equation}

To calculate how the critical force depends on the wavenumber $q$, that characterizes the mode of deformation, we truncate the polynomial given by Eq.~\ref{SI:polinom} to first order in $\mu$ and solve it for $\mu$. We obtain
\begin{equation}
    \label{SI:truncated_polynom}
    \mu= \frac{\alpha}{\beta}\>,
\end{equation}
where
\begin{equation}
    \begin{aligned}
    \alpha&=\sqrt{\Gamma } K [128 \Gamma ^3+q^6 \left(5 \Gamma ^2+8 \Gamma  \Delta -8\right)+8 \Gamma  \left(\Gamma ^2+5\right) q^4+\\
    &+16 \Gamma ^2 q^2 \left(\Gamma ^2+2 \Gamma  \Delta -2\right)]+4 \Gamma ^2 q^4 [\Gamma ^2 (2 B +1) q^2+\\
    &+4 \Gamma  \left(-\Delta ^2+4 B +\Delta  B  q^2+2\right)-q^2 \left(\Delta ^2+4 B \right)] \quad
    \end{aligned}
\end{equation}
and
\begin{equation}
    \begin{aligned}
    \beta&=2 \Gamma ^{3/2} q^2 [32 \Gamma ^2+q^4 \left(-3 \Gamma ^2-4 \Gamma  \Delta  B +\Delta ^2+8 B +14\right)+\\
    &+4 \Gamma  q^2 \left(\Gamma ^2+\Delta ^2-8 B -12\right)]-\\
    &-2 K \left(64 \Gamma ^3+2 q^6 (\Gamma  \Delta -2)+5 \Gamma  \left(\Gamma ^2+4\right) q^4+8 \Gamma ^2 q^2 [\Gamma  \Delta -2]\right).
    \end{aligned}
\end{equation}
A Taylor expansion of $\mu$ around $q_0 = 0$ for small $q$ diverges unless $K$ is set to zero (i.e., in the absence of stroma). This indicates that the divergence arises because the small-$q$ limit inherently corresponds to small values of $K$. This can be solved by assuming a scaling between $K$ and $q$. We assume that this scaling is the same as in supported solid plates, where $q\propto K^{1/3}$. We then neglect all terms in $\beta$ that are proportional to $q^n$ with $n\geq 3$. Simplifying Eq.~\ref{SI:truncated_polynom} in this manner and expending it in terms of powers of $q$, we obtain an in-plane force, which reads
\begin{equation}
    \mu(q) = \frac{K}{q}+\frac{q K \left(h^4+2 h^2  \Delta -2\right)}{8 h^2 } + \frac{q^2\left(8B-\Delta^2+2\right)}{8}+ \frac{q^3 K \left(h^4+5\right) }{16 h^4} +\frac{q^4 \left(4 B h^4+8 B h^2  \Delta -8 B+h^4-\Delta ^2\right)}{32 h^2 }.
    \label{SI:critical_mu}
\end{equation}
Here and hereon, we replace $\Gamma$ in all equations by $h^2$ (see Eq.~\ref{SI:hh}).

\section{Elastic instability}
At the critical point, the flat configuration looses stability and either buckles or wrinkles out of plane. The deformation mode is given by the optimal wave-number $q_0$, obtained by
\begin{equation}
    \frac{\partial \mu}{\partial q}\Big\vert_{q=q_0} = 0\>.
    \label{SI:polynom}
\end{equation}
\subsection{No apico-basal tension polarity}
First we set to investigate a minimal system where $K=B=\Delta=0$ and the only model parameter with a non-zero value is $\Gamma$. Equation~\ref{SI:critical_mu} simplifies to 
\begin{equation}
    \mu(q)=\frac{ q^2}{4}\>,
\end{equation}
such that $\mu(q)$ is minimized at $q=0$, meaning that the tissue either stays flat and uniformly increases it's height or buckles out of plane.

At the onset of buckling transition $q=2\pi/L$, where $L$ is the end-to-end length of the tissue and the critical force and the critical compressive strain read
\begin{equation}
    \mu_0=\frac{ \pi^2}{L^2} \,\>.
\end{equation}
and
\begin{equation}
    \epsilon_0=\frac{\pi^2}{2h^2 L^2}\>,
\end{equation}
respectively. 

\subsection{Basement membrane}
In absence of the stroma and the differential tension ($K=\Delta=0$) but the basement membrane present ($B>0$), Eq.~\ref{SI:critical_mu} simplifies to
\begin{equation}
    \mu^{[B]}(q) =  \frac{q^2\left(4B+1\right)}{4} \;\>
    \label{SI:mu_B}
\end{equation}
and, again, $\mu(q)$ is minimized at $q=0$, implying buckling. The corresponding critical compressive strain
\begin{equation}
    \epsilon_c^{[B]}= \frac{\mu^{[B]}\left(q=2\pi/L\right)}{2h^2}=\epsilon_0 \left(1+4B\right)=\frac{\pi ^2 \left(1+4B\right)}{2 h^2 L^2}\;\>.
\end{equation}

\subsection{Differential tension}
In absence of the stroma and the basement membrane ($K=B=0$) but the differential tension present ($\Delta\neq 0$) Eq.~\ref{SI:critical_mu} simplifies to
\begin{equation}
    \mu^{[\Delta]}(q) = \frac{q^2\left(2-\Delta^2\right)}{8}+\frac{q^4 \left(h^4-\Delta ^2\right)}{32 h^2 }\; .
    \label{SI:mu_delta}
\end{equation}
While the $q^2$-term in Eq.~\ref{SI:mu_delta} is sufficient to calculate the critical compressive strain, the $q^4$-term is needed to determine the optimal wave-vector.

When $\left |\Delta\right |<\sqrt{2}$, $\mu^{[\Delta]}(q)$ is minimized at $q=0$ and the tissue of length $L$ buckles with a wave-number $q=2\pi/L$ once it is compressed beyond the critical compressive strain
\begin{equation}
    \epsilon_{\rm buck}^{[\Delta]}= \frac{\mu^{[\Delta]}\left(q=2\pi/L\right)}{2h^2}=\epsilon_0 \left(1-\frac{\Delta^2}{2}\right)=\frac{\pi ^2 \left(2-\Delta ^2\right)}{4 h^2L^2}\;\>.
\end{equation}

Once the value of $|\Delta|$ reaches a critical value $\Delta_c=\sqrt{2}$, the critical compressive strain $\epsilon_{\rm buck}^{[\Delta]}=0$, while the optimal wave-vector, determined from Eq.~\ref{SI:mu_delta} and expanded to the first order around $|\Delta|=\sqrt{2}$, becomes 
\begin{equation}
    q_0^{[\Delta]} =\frac{2 h \sqrt{\Delta ^2-2}}{\sqrt{\left(h^4-8\right) \Delta ^2+12}} \approx \dfrac{2^{5/4}\left (\left|\Delta\right |-\sqrt{2}\right )^{1/2}}{\sqrt{h^2-2h^{-2}}}\>,
    \label{SI:q0_delta}
\end{equation}
corresponding to a wrinkling instability. Here, the reference flat tissue is unstable, which can be seen from the fact that the critical compressive strain becomes negative:
\begin{equation}
    \epsilon_{\rm wrink}^{[\Delta]}=\frac{\mu^{[\Delta]}\left(q=q_0^{[\Delta]}\right)}{2h^2}=-\frac{\left(\Delta -\sqrt{2}\right)^2}{2 \left(h^4-2\right)}\,\>,
\end{equation}
where only the first term in Eq.~\ref{SI:mu_delta} was included. 

\subsection{Differential tension and basement membrane}
In absence of the stroma ($K=0$), the effects of the differential tension and the basement membrane appear at the same order in Eq. \ref{SI:critical_mu}. The critical force for harmonic deformation reads
\begin{equation}
    \mu^{[\Delta, B]}(q) = \frac{q^2\left(8B-\Delta^2+2\right)}{8}+\frac{q^4 \left(4 B h ^4+8 B h^2  \Delta -8 B+h^4-\Delta ^2\right)}{32 h^2 }.
    \label{SI:mu_delta_b}
\end{equation}
The competing effects of basement membrane and the differential tension, which prefer flat and curved configurations, respectively, result in either buckling or wrinkling deformation mode with bending modulus $B$ shifting the critical value of $|\Delta|$ to
\begin{equation}
    \Delta_c=\sqrt{2 (1+4B)}\>.
\end{equation}
For $|\Delta|< \sqrt{2 (1+4B)}$ buckling is preferred (i.e., $\mu^{[\Delta,B]}(q)$ is minimized at $q=0$) and occurs at a critical compressive strain
\begin{equation}
    \epsilon_{\rm buck}^{[\Delta, B]}=\frac{\mu^{[\Delta, B]}(q=2\pi/L)}{2h^2}=\epsilon_0\left(1-\frac{\Delta^2}{2} + 4B\right)\; . 
\end{equation}
The optimal wave-number in the wrinkling regime, expanded to the first non-zero order around $\Delta=\Delta_c$, reads
\begin{equation}
    q_0^{[\Delta, B]} =\frac{2 h \sqrt{-8 B+\Delta ^2-2}}{\sqrt{64 B^2+8 B \left(2 h^2  \Delta -\Delta ^2+6\right)+\left(h^4-8\right) \Delta ^2+12}} \approx \dfrac{2^{5/4}\left (\left|\Delta\right |-\sqrt{2(1+4B)}\right )^{1/2}}{\sqrt{h^2-2/h^2}}\>.
    \label{SI:q0_D_B}
\end{equation}
The critical compressive strain in the wrinkling regime, obtained by considering only the first term in Eq.~\ref{SI:mu_delta_b}, reads
\begin{equation}
    \epsilon_{\rm wrink}^{[\Delta, B]}=\frac{\mu^{[\Delta, B]}\left(q=q_0^{[\Delta, B]}\right)}{2h^2}=-\frac{\left(1+4B\right)\left(\Delta -\sqrt{2(1+4B)}\right)^2}{2 \left(h ^4-2\right)}\, .
\end{equation}

\subsection{Stroma}
In absence of the differential tension and the basement membrane ($\Delta=B=0$), the equation \ref{SI:critical_mu} simplifies into
\begin{equation}
    \mu^{[K]}(q) = \frac{K}{q}+\frac{q K \left(h^4 -2\right)}{8 h^2 } + \frac{q^2}{4},
    \label{SI:mu_K}
\end{equation}
where the lowest-order terms in $q$, necessary to obtain the solution are kept. When $K>0$, the wrinkling instability appears with the optimal wave-number
\begin{equation}
    q_0^{[K]}=\frac{1}{12 \Gamma }\left[\eta^{[K]}+ \frac{\left(\Gamma ^2-2\right)^2 K^2}{\eta^{[K]}}-\left(\Gamma ^2-2\right) K \right ]\>,
\end{equation}
where
\begin{equation}
    \eta^{[K]} = \sqrt[3]{-\left(\Gamma ^2-2\right)^3 K^3+24 \sqrt{6} \sqrt{864 \Gamma ^6 K^2-\Gamma ^3 \left(\Gamma ^2-2\right)^3 K^4}+1728 \Gamma ^3 K}\: .
\end{equation}
To the lowest order in $K$,
\begin{equation}
    q_0^{[K]}\approx \left(2 K \right)^{1/3}\;\>.
    \label{SI:q_K}
\end{equation}
We obtain the critical compressive strain, necessary to trigger wrinkling deformation, by considering only the first term of Eq. \ref{SI:mu_K} and truncated wave vector $q_0$ \ref{SI:q_K} as
\begin{equation}
    \epsilon_c^{[K]} =\frac{\mu^{[K]}\left(q=q_0^{[K]}\right)}{2h^2}=3 K^{2/3}2^{-7/3}  h^{-2}\; .
\end{equation}

\subsection{Stroma, differential tension, and basement membrane}
The general case, when ($\Delta\neq0$, $B\neq 0$, and $K\neq 0$, requires solving a fifth-degree polynomial in $q$ (Eq.~\ref{SI:critical_mu}). To avoid this, we assume that the primary mechanism of wrinkling is the elastic interaction with the stroma and we consider the effects by the differential tension and the basement membrane perturbatively. In particular, we write 
\begin{equation}
    q^{[K, \Delta, B]}_0=q_0^{[K]} + \delta q(K, \Delta, B)\; ,
    \label{SI:ansatz_K_D_B}
\end{equation}
assuming $\delta q \ll q_0^{[K]}$. Inserting this {\it ansatz} into Eq.~\ref{SI:critical_mu} and solving Eq.~\ref{SI:polynom} 
%
%
yields, to the linear order in $\delta q$,
\begin{equation}
    \begin{aligned}
     \delta q &\left[2 K \left(32 B \left[h^4+2 h^2  \Delta -2\right]+9 h^4+2 h^2  \Delta -8 \Delta ^2-2\right)+ 2^{10/3} K^{1/3} h^2   \left(8 B-\Delta ^2+3\right)+9\ 2^{2/3} \left(h^4+5\right) K^{5/3} h^{-2}\right]+\\
     &+\left[2 \sqrt[3]{2} K^{4/3} \left(8 B \left[h^4+2 h^2  \Delta -2\right]+3 h^4+2 h^2  \Delta -2 \Delta ^2-2\right)+4\ 2^{2/3} h^2  K^{2/3} \left(8 B-\Delta ^2\right)+6 \left(h^4+5\right) K^2 h^{-2}\right]=0\>.
    \end{aligned}
    \label{SI:longEq4dq}
\end{equation}
By solving Eq.~\ref{SI:longEq4dq} for $\delta q$ and truncating the solution firstly to the lowest order in $K$ and secondly to the second order in both $\Delta$ and $B$, we obtain
\begin{equation}
    \delta q = q_0^{[K]}\ \frac{1}{6}\left(\Delta ^2-8 B \right) + \left(q_0^{[K]}\right)^3 \frac{\Delta(1+8B)}{12}\;
\end{equation}
and the optimal wave-number reads
\begin{equation}
    q_0^{[K, \Delta, B]}=q_0^{[K]} \left(1+ \frac{1}{6}\left[\Delta^2 - 8B\right] \right)\;.
    \label{SI:q0_K_D_B}
\end{equation}
Note that due to approximations assumed in the derivation, formula is not valid in the regime $|\Delta| \gg \sqrt{2}$, where wrinkling is primarily caused by the differential tension.

\section{Thickness modulation}
Assuming the solution in the form
\begin{equation}
    \begin{aligned}
    l(\sigma)&=\delta h + \Lambda \cos(q \sigma)\; ,\\
    \psi(\sigma) &= \Psi \sin(q \sigma)\; , \\
    y(\sigma) &=Y \cos(q \sigma)\>,
    \end{aligned}
\end{equation}
we can quantify the groove-to-crest thickness modulation by defining
\begin{equation}
    \tau=\frac{\Lambda}{Y},
\end{equation}
The phase of thickness modulation relative to substrate undulations is determined through the sign of $\tau$ -- a positive sign means that the two are in phase, while a negative sign implies that the phases are shifted by $\pi/2$ (anti-phase).

As the variable $y(\sigma)$ is eliminated early in solving E-L equations, we express the variable $Y$ through the constraint, given by Eq.~\ref{SI:stroma_constraint}, which yields
\begin{equation}
    Y=-\frac{\Lambda}{2} - \frac{\Psi}{q}\>.
\end{equation}
Thus, the quantity $\tau$ is then expressed as
\begin{equation}
    \tau=\frac{\Lambda}{Y} =\frac{2\Lambda}{-\Lambda-2\Psi/q} = \frac{-q\Lambda/\Psi}{1+q\Lambda/(2\Psi)} \approx  -\frac{q\Lambda}{\Psi}\; ,
\end{equation}
assuming $q \ll 1$ and $\Lambda \sim \Psi$. The ratio $\Lambda/\Psi$ can be directly expressed from Eq.~\ref{SI:EL_equation_1} for every combination of non-zero values of parameters $K$, $\Delta$, and $B$. 

\subsection{No apico-basal tension polarity}
In absence of the differential tension and the substrate ($K=\Delta=B=0$), Eq.~\ref{SI:EL_equation_1} can be solved using
\begin{align}
    l&=\delta h +\Lambda\exp( i q \sigma)\; ,\\
    \dot Q &= 0\>,
\end{align}
which gives
\begin{equation}
    32 h^4 \delta h (h^2 +\mu )-16 h^5 \mu +\Lambda \exp(i q \sigma) \left[32 h^4(h^2 +\mu )+q^4 \left(h^6+2 h^2 +2 \mu \right)+4 h^2  q^2 \left(h^6-2 h^2 -2 \mu \right)\right]=0\>.
\end{equation}
The sum of first two terms must equal to zero, which gives the relation
\begin{equation}
    \delta h = \frac{h \mu }{2 (h^2 +\mu )} \approx \frac{\mu}{2 h} \quad \implies \quad \mu = 2 h\delta h\; .
    \label{SI:dl0_gamma}
\end{equation}
The last term must also give zero, which can be satisfied by either of the two conditions:
\begin{equation*}
    \begin{aligned}
    (a)\: &\Lambda=0\quad {\rm or}\\
    (b)\: &\mu= \frac{\Gamma  \left(32 \Gamma ^2+\left(\Gamma ^2+2\right) q^4+4 \Gamma  \left(\Gamma ^2-2\right) q^2\right)}{2 \left(16 \Gamma ^2+q^4-4 \Gamma  q^2\right)} \overset{\ q\ll 1}{\approx}-h^2-\frac{1}{8} \left(h^4 q^2\right)\>.
    \end{aligned}
\end{equation*}
The condition (b) in not physical and contradicts the necessary equality between $\mu$ and $\delta h$ in Eq.~\ref{SI:dl0_gamma}. The only possible solution is then $\Lambda=0$ and thus,
\begin{equation}
    \tau_0 =\frac{\Lambda}{Y}=0\>,
\end{equation}
meaning that the shape of crests and grooves are only mirrored images of each other, as expected from the apico-basal symmetry.

\subsection{Basement membrane}
In the absence of the differential tension and the stroma ($\Delta=K=0$) but the basement membrane present ($B>0$), we use the following {\it ansatz}
\begin{align}
    l&=\delta h +\Lambda\cos(q \sigma)\; ,\\
    \psi&=\Psi \sin(q \sigma)\; , \\
    \dot Q &= 0
\end{align}
to solve Eq.~\ref{SI:EL_equation_1}, which then reads
\begin{multline}
    32 \Gamma ^2 \delta h (\Gamma +\mu )-16 \Gamma ^{5/2} \mu + \\
    +\cos (q \sigma) \left[\Lambda \left(32 \Gamma ^2 [\Gamma +\mu ]+q^4 \left[4 \Gamma ^3 B +\Gamma ^3+2 \Gamma +2 \mu \right]+4 \Gamma  q^2 \left[\Gamma ^3-2 \Gamma -2 \mu \right]\right)+8 \Gamma ^3 B  q^3 \Psi \right]=0\;.
\end{multline}
Here, the equality is satisfied when the sum of first two terms equals to zero:
\begin{equation}
    \delta h = \frac{h \mu }{2 (h^2 +\mu )} \approx \frac{\mu}{2 h} \quad \implies \quad \mu = 2 h\delta h\>
\end{equation}
and the last term is zero as well. The latter is true when 
\begin{equation}
    \frac{\Lambda}{\Psi}=-\frac{8 \Gamma ^3 B  q^3}{32 \Gamma ^2 (\Gamma +\mu )+q^4 \left[\Gamma ^3 (4 B +1)+2 \Gamma +2 \mu \right]+4 \Gamma  q^2 \left(\Gamma ^3-2 \Gamma -2 \mu \right)}\; .
    \label{SI:l1_psi0_B}
\end{equation}
By using the relation between $\mu$ and $q$ (Eq.~\ref{SI:mu_B})  and then expanding Eq.~\ref{SI:l1_psi0_B} to the smallest order in $q$, we obtain
\begin{equation}
    \frac{\Lambda}{\Psi}\approx -\frac{q^3\,B}{4}.
\end{equation}
Thus, the thickness modulation in a buckled tissue of length $L$, supported by a basement membrane reads
\begin{equation}
    \tau^{[B]} = \frac{q^4\,B}{4} = \frac{4B\pi^4}{L^4}>0\>.
\end{equation}

\subsection{Differential tension}
To analyze the thickness modulation in tissues where the substrate interaction is absent ($K=B=0$) but cells are apico-basally polarized ($\Delta\neq 0$), we first observe Eq.~\ref{SI:EL_equation_1}, assuming $\dot Q=0$:
\begin{equation}
    4 \Gamma ^{5/2} \left(\sqrt{\Gamma } \left(\Gamma  \ddot l+2 \Delta \dot \psi \right)+4 \mu +\Delta  \ddot \psi \right)=\left(\Gamma ^3+2 \Gamma +2 \mu \right) l^{(4)}+8 \Gamma  (\Gamma +\mu ) \ddot l+32 \Gamma ^2 (\Gamma +\mu ) l+\Gamma ^{3/2} \Delta  \psi ^{(4)}.
\end{equation}
Because this equation contains both odd and even derivatives of $\psi$, we use the following {\it ansatz}:
\begin{align}
    l&=\delta h + \Lambda \cos( q \sigma)\; ,\\
    \psi&=\Psi \sin( q \sigma) + \Psi_1 \cos(q \sigma)\>
\end{align}
and obtain
\begin{multline}
    \Gamma ^{3/2} \left[-16 \Gamma  \mu +32 \sqrt{\Gamma } \delta h (\Gamma +\mu )\right]+\Gamma ^{3/2}\Delta  q \sin (q s) \left(8 \Gamma ^{3/2} \Psi_1+q^3 \Psi+4 \Gamma  q \Psi\right)+\\
    +\cos (q s) \left[\Lambda \left(32 \Gamma ^2 [\Gamma +\mu ]+q^4 \left[\Gamma ^3+2 \Gamma +2 \mu \right]+4 \Gamma  q^2 \left[\Gamma ^3-2 \Gamma -2 \mu \right]\right)+\Gamma ^{3/2} \Delta  q \left(-8 \Gamma ^{3/2} \Psi+q^3 \Psi+4 \Gamma  q \Psi_1\right)\right]=0\; .
    \label{SI:polynom_tau_delta}
\end{multline}
Setting the first three terms in Eq.~\ref{SI:polynom_tau_delta} to zero, yields
\begin{equation}
    \begin{aligned}
    &(a)\; \delta h = \frac{h \mu }{2 (h^2 +\mu )} \approx \frac{\mu}{2 h} \quad \implies \quad \mu = 2 h\, \delta h\; ,\\
    &(b)\; \Psi_1= -q \Psi\, \frac{ 4 \Gamma +q^2}{8 \Gamma ^{3/2}}\; ,\\
    &(c)\; \Lambda = -\frac{\Gamma ^{3/2} \Delta  q \left(-8 \Gamma ^{3/2} \Psi+q^3 \Psi_1+4 \Gamma  q \Psi_1\right)}{32 \Gamma ^2 (\Gamma +\mu )+q^4 \left(\Gamma ^3+2 \Gamma +2 \mu \right)+4 \Gamma  q^2 \left(\Gamma ^3-2 \Gamma -2 \mu \right)}\; .
    \label{SI:EL1_delta}
    \end{aligned}
\end{equation}
We use the first term of the expression for the critical force from (Eq.~\ref{SI:mu_delta}) and the equality $(b)$ from Eq.~\ref{SI:EL1_delta} to simplify the expression $(c)$ from Eq.~\ref{SI:EL1_delta} to
\begin{equation}
    \frac{\Lambda}{\Psi}=-\frac{\Delta  q  \left(64 \Gamma ^3+q^6+8 \Gamma  q^4+16 \Gamma ^2 q^2\right)}{-256 \Gamma ^3+2 \left(\Delta ^2-2\right) q^6-8 \Gamma  q^4 \left(\Gamma ^2+\Delta ^2\right)-32 q^2 \left(\Gamma ^4-\Gamma ^2 \Delta ^2\right)}\approx \frac{q\, \Delta}{4}\>
\end{equation}
and thus,
\begin{equation}
    \tau^{[\Delta]}=-\frac{q^2\, \Delta}{4}\; .
\end{equation}
In the buckling regime,
\begin{equation}
    \tau^{[\Delta]}_{\rm buck}=\tau^{[\Delta]}\left(q=\frac{2\pi}{L}\right)=-\frac{\pi^2 \Delta}{L^2}\>,
\end{equation}
whereas in the wrinkling regime,
\begin{equation}
    \tau^{[\Delta]}_{\rm wrink}=\tau^{[\Delta]}\left(q=q_0^{[\Delta]}\right)= -\frac{\sqrt{2}\Delta \left(|\Delta| -\sqrt{2}\right)}{h^2-2/h^2}\; .
\end{equation}
\subsection{ Differential tension and basal membrane}
To analyze the thickness modulation in tissues where the substrate interaction is absent ($K=0$) but cells are apico-basally polarized and supported by basement membrane (i.e., $\Delta, B\neq 0$), we first observe Eq.~\ref{SI:EL_equation_1}, assuming $\dot Q=0$:
\begin{equation}
    \begin{aligned}
    &4 \Gamma ^{5/2} \left(\sqrt{\Gamma } \left(\Gamma  \ddot l+2 \Delta  \dot \psi \right)+4 \mu +\Delta  \ddot \psi \right)+8 \Gamma ^3 B  \psi ^{(3)}= \\ &=l^{(4)} \left(4 \Gamma ^3 B +\Gamma ^3+2 \Gamma +2 \mu \right)+8 \Gamma  (\Gamma +\mu )\ddot l+32 \Gamma ^2 (\Gamma +\mu ) l+\Gamma ^{3/2} \Delta  \psi ^{(4)}.
    \end{aligned}
\end{equation}
Similarly to the system with only differential tension present, this equation contains both odd and even derivatives of $\psi$, and we use the following {\it ansatz}:
\begin{align}
    l&=\delta h + \Lambda \cos( q \sigma)\; ,\\
    \psi&=\Psi \sin( q \sigma) + \Psi_1 \cos(q \sigma)\>
\end{align}
which yields the equation
\begin{equation}
    \begin{aligned}
        &\Gamma ^{3/2} \left[-16 \Gamma  \mu +32 \sqrt{\Gamma } \delta h (\Gamma +\mu )\right]+q \sin (q s) \Gamma ^{3/2}\left[8 \Gamma ^{3/2} \Psi_1 \left(\Delta -\kappa  q^2\right)+\Delta  q \Psi \left(4 \Gamma +q^2\right)\right]+\\ 
        &+\cos (q s) [\Lambda \left(32 \Gamma ^2 (\Gamma +\mu )+q^4 \left(\Gamma ^3 (4 \kappa +1)+2 \Gamma +2 \mu \right)+4 \Gamma  q^2 \left(\Gamma ^3-2 \Gamma -2 \mu \right)\right)+\\ &+\Gamma ^{3/2} \Delta  q^2 \Psi_1 \left(4 \Gamma +q^2\right)+8 \Gamma ^3 q \Psi \left(\kappa  q^2-\Delta \right)]=0
    \end{aligned}
\end{equation}
Setting all three terms in above expression to 0 yields
\begin{equation}
    \begin{aligned}
        &(a)\; \delta h = \frac{h \mu }{2 (h^2 +\mu )} \approx \frac{\mu}{2 h} \quad \implies \quad \mu = 2 h\, \delta h\; ,\\
        &(b)\; \Psi_1=-\frac{\Delta  q \Psi \left(4 \Gamma +q^2\right)}{8 \Gamma ^{3/2} \left(\Delta -B  q^2\right)}\; ,\\
        &(c)\; \Lambda = -\frac{\Gamma ^{3/2} q \left(\Delta  q^3 \Psi_1-8 \Gamma ^{3/2} \Psi \left(\Delta -B  q^2\right)+4 \Gamma  \Delta  q \Psi_1\right)}{32 \Gamma ^2 (\Gamma +\mu )+q^4 \left(\Gamma ^3 (4 B +1)+2 \Gamma +2 \mu \right)+4 \Gamma  q^2 \left(\Gamma ^3-2 \Gamma -2 \mu \right)}\; .
        \label{SI:EL1_delta_B}
    \end{aligned}
\end{equation}
We use the first term in expression \ref{SI:mu_delta_b} and relation $(b)$ from solution \ref{SI:EL1_delta_B} to replace $\mu$ and $\Psi_1$, respectively, in expression \ref{SI:EL1_delta_B}$(c)$. The amplitude $\Lambda$, truncated to third order in $q$, thus becomes
\begin{equation}
    \Lambda=\Psi \frac{\Delta  q }{4} + \Psi \frac{q^3  \left(\Delta  \left(-\Gamma ^{5/2}+4 \sqrt{\Gamma }+\Delta ^2-2\right)-8 B \left(\Gamma ^{3/2}+\Delta \right)\right)}{32 \Gamma ^{3/2}}\>.
\end{equation}
In the buckling regime ($|\Delta|<\sqrt{2(1+4B)}$) a tissue of length $L$ with both differential tension and basement membrane present will exhibit thickness modulation
\begin{equation}
    \tau^{[\Delta, B]}_{\rm buck} = -\frac{\Delta \pi^2}{L^2} + \frac{8 \pi ^4 B \left(\Gamma ^{3/2}+\Delta \right)+\pi ^4 \Delta  \left(\Gamma ^{5/2}-4 \sqrt{\Gamma }-\Delta ^2+2\right)}{2 \Gamma ^{3/2} L^4}\>.
    \label{SI:tau_D_B_buck}
\end{equation}
In the wrinkling regime ($|\Delta|>\sqrt{2(1+4B)}$) with wave-vector $q^{[\Delta, B]}$ \ref{SI:q0_D_B} the fist term in \ref{SI:tau_D_B_buck} suffices to observe the effects of both $\Delta$ and $B$ on the tissue thickness modulation as
\begin{equation}
    \tau^{[\Delta, B]}_{\rm wrink} = -\frac{  \Delta \sqrt{8 B+2} \left(\Delta -\sqrt{2+8B}\right)}{\Gamma -2/\Gamma}\>.
\end{equation}

\subsection{Stroma}
For a tissue without any apico-basal tension polarity ($\Delta=0$), attached to a stroma ($K>0$) but not to a basement membrane ($B=0$), Eq.~\ref{SI:EL_equation_1} is first differentiated with respect to $\sigma$ and the last E-L equation \ref{SI:EL_equation_4} is used to replace variable $\ddot Q$. The resulting equation reads
\begin{equation}
    \begin{aligned}
    &4 \Gamma ^2 \left[\frac{1}{4}\Gamma^{-3/2}  K \left(\psi ^{(4)} -\frac{1}{2} l^{(5)} \right)+2 \Gamma ^{1/2} K\left(\psi  -\frac{\dot l }{2}\right) \Gamma^{3/2}  l^{(3)} \right]=\\
    &=l^{(5)}  \left[\sqrt{\Gamma } \left(\Gamma ^2+2\right)+2 \mu \right]+8 \Gamma  \left(\sqrt{\Gamma }+\mu \right) l^{(3)} +32 \Gamma ^2 \left(\sqrt{\Gamma }+\mu \right) \dot l \; . 
    \end{aligned}
\end{equation}
As we will only be interested in solutions close to the onset of wrinkling where $q \ll 1$, we neglect fourth and higher derivatives. We use the following {\it ansatz}:
\begin{equation}
    \begin{aligned}
    l&=\delta h + \Lambda \cos(q \sigma)\; ,\\
    \psi&=\Psi \sin(q \sigma)\;  
    \end{aligned}
\end{equation}
and the expression for the critical force given by Eq.~\ref{SI:mu_K} to obtain
\begin{equation}
    \frac{\Lambda}{\Psi}=\frac{8 \Gamma ^3 K q }{K \left(-32 \Gamma ^2+\left(\Gamma ^2-2\right) q^4-8 \Gamma  \left(\Gamma ^2-2\right) q^2\right)+2 \Gamma  q \left(-16 \Gamma ^2+q^4-2 \Gamma ^3 q^2\right)}\;.
    \label{SI:l1_k}
\end{equation}
We use the solution for the optimal wave-number $q_0^{[K]}=(2K)^{1/3}$ to replace $q$ in Eq.~\ref{SI:l1_k} and then expand it to the first order in $K$ around $K=0$. We obtain
\begin{equation}
   \frac{\Lambda}{\Psi}\approx - \frac{K}{4}
\end{equation}
and, finally,
\begin{equation}
    \tau^{[K]} = 2^{-5/3}\, K^{4/3}\;.
\end{equation}

\subsection{Stroma, differential tension, and basement membrane}
To solve a general case where $K \neq0, \Delta \neq 0,$ and $B\neq 0$ we return to the complete Eq. \ref{SI:EL_equation_1} and use the $ansatz$
\begin{align*}
    l(\sigma)&=\delta h + \Lambda \cos(q \sigma)\; ,\\
    \psi(\sigma) &= \Psi \sin(q \sigma) + \Psi_1 \cos(q \sigma)\;.  
\end{align*}
The Eq. \ref{SI:EL_equation_1} becomes {}
\begin{multline}
    \frac{1}{4 \Gamma ^{3/2}}K q \left[8 \Gamma ^2 \left(\frac{1}{2} \Psi q \sin (q \sigma)+\Psi \sin (q \sigma)+\Psi_1 \cos (q \sigma)\right)+q^4 \Psi \sin (q \sigma)+q^4 \Psi_1 \cos (q \sigma)\right]+ \\
    +\sqrt{\Gamma } \left[\Gamma  \Lambda q^3 \sin (q \sigma)+2 \kappa  \left(q^4 \Psi \sin (q \sigma)+q^4 \Psi_1 \cos (q \sigma)\right)\right]+\Delta  \left[q^3 \Psi_1 \sin (q \sigma)-q^3 \Psi \cos (q \sigma)\right] + \\ 
    +2 \sqrt{\Gamma } \Delta  \left[q^2 (-\Psi) \sin (q \sigma)-q^2 \Psi_1 \cos (q \sigma)\right]-\frac{1}{4 \Gamma^2}\left(8 \Gamma  \Lambda q^3 \left(\sqrt{\Gamma }+\mu \right) \sin (q \sigma)-32 \Gamma ^2 \Lambda q \left(\sqrt{\Gamma }+\mu \right) \sin (q \sigma)\right)=0\;,
\end{multline}
which we solve to obtain thickness amplitude as
\begin{multline}
    \Lambda = -\frac{\cot (q s) \left[8 B \Gamma ^2 q^3 \Psi_1+8 \Gamma ^2 K \Psi_1+K q^4 \Psi_1-4 \Gamma ^{3/2} \Delta  q^2 \Psi-8 \Gamma ^2 \Delta  q \Psi_1\right]}{4 \left(8 \left(\Gamma ^{3/2} \mu +\Gamma ^2\right)+\Gamma ^2 K q+q^2 \left(\Gamma ^3-2 \sqrt{\Gamma } \mu -2 \Gamma \right)\right)}-\\
    -\frac{8 B \Gamma ^2 q^3 \Psi+K \Psi \left(8 \Gamma ^2+q^4\right)+4 \Gamma ^{3/2} \Delta  q^2 \Psi_1-8 \Gamma ^2 \Delta  q \Psi}{4 \left(8 \left(\Gamma ^{3/2} \mu +\Gamma ^2\right)+\Gamma ^2 K q+q^2 \left(\Gamma ^3-2 \sqrt{\Gamma } \mu -2 \Gamma \right)\right)}\;.
    \label{SI:L_K_D_B}
\end{multline}
We solve for amplitude $\Psi_1$ by setting the first (non-constant) term in the above expression to 0 and obtain 
\begin{equation}
    \Psi_1=\frac{4 \Gamma ^{3/2} \Delta  q^2 \Psi}{K \left(8 \Gamma ^2+q^4\right)+8 \Gamma ^2 q \left(B  q^2-\Delta \right)}\;.
\end{equation}
Further on we simplify the Eq. \ref{SI:L_K_D_B} for thickness amplitude by the expression \ref{SI:q0_K_D_B} for optimal wave-vector and expression \ref{SI:critical_mu} for critical force and finally truncation it for small values of  constants $K, \Delta,$ and $B$. The two lowest orders read 
\begin{equation}
    \frac{\Lambda}{\Psi}^{[K, \Delta, B]}= -\frac{K^{1/3} (4 B-3) \Delta  }{6\ 2^{2/3}} - \frac{1}{4}K 
\end{equation}
and the thickness modulation is obtained as
\begin{equation}
    \tau^{[K, \Delta, B]}=\frac{ K^{4/3}}{ 2^{5/3}}\left(1-\frac{4}{3}B\right)-\frac{K^{2/3} \left((3-4 B)^2 \Delta  \right)}{2^{1/3}\; 18} =\tau^{[K]}\left[  1  - \frac{4}{3}B - \frac{2^{1/3} \Delta}{K^{2/3}}\left(1-\frac{4}{3}B\right)^2\right]\>. 
\end{equation}
Thickness modulation changes sign as $\Delta$ reaches
\begin{equation}
    \Delta_i^{[\Delta,B,K]} =  \frac{\left(2K\right)^{2/3}}{ 2(1-4B/3)}\>.
\end{equation}

\section{Variation of thickness through the variation of cell cross-section area}
In addition to varying cell aspect ratio at a fixed cell cross-section area, cell height can also be varied by varying cell cross-section area and keeping the aspect ratio fixed. Here, we investigate the effect of the latter on our main results.

We begin by determining the reference energetically optimal shape of a quadrilateral cell. The overall energy is normalized by $\Gamma_l \sqrt{A_0}$, where $A_0$ is a reference cell cross-section area. The actual cell area $A$ in dimensionless form is thus given in the units of $A_0$: $a = A/A_0$. Minimizing the energy of a flat epithelium yields the optimal cell height $H$ and width $D$ as
\begin{align*}
    H&=\sqrt{\Gamma \, a}\;,\\
    D&=\frac{a}{H} = \sqrt{\frac{a}{\Gamma}}\;.
\end{align*}
We repeat the derivation of the elasticity theory, where, compared to the above derivation, three steps need to be adjusted:
\begin{enumerate}
    \item cell midlines change to
    \begin{equation}
        s_1=\frac{2 a}{l_1 \cos (\phi_1-\psi_1)+l_2 \cos (\phi_2-\psi_1)} \quad \textrm{and} \quad s_2=\frac{2 a}{l_2 \cos (\phi_2-\psi_2)+l_3 \cos (\phi_3-\psi_2)} \>,
    \end{equation}
    \item the cell height and width of cells in the reference state change to $H$ and $D$, respectively,
    \item the discrete step $1/h$ is replaced by $D$ when rewriting the discrete variables as a Taylor expansion of continuum variables.
\end{enumerate}
The system of Euler-Lagrange equation reads
\begin{equation}
    \begin{aligned}
        &\Gamma  \Delta  \sqrt{a^7 \Gamma } \psi ^{(4)}+a^2 l^{(4)} \left(\Gamma ^3 \left(a^2+4 B\right)+2 a \mu +2 \Gamma \right)+16 \Gamma ^2 (a \Gamma )^{3/2}+8 \Gamma  (a \mu +\Gamma ) \left(a \ddot l+4 \Gamma  l\right)= \\
        &=4 \left(a^{3/2} \Gamma ^{7/2} \dot Q+a^3 \Gamma ^4 \ddot l+2 a^2 \Gamma ^3 \left(B \psi ^{(3)}+\Delta  \dot \psi \right)+4 \sqrt{a \Gamma ^7}+4 a \mu  \sqrt{a \Gamma ^5}+\Delta  (a \Gamma )^{5/2} \ddot \psi \right), \\
        & \Gamma  (a \mu +\Gamma ) \left(a \left(-a \psi ^{(4)}+2 a \phi ^{(4)}-4 \Gamma  \ddot \psi \right)+16 \Gamma ^2 \phi \right)=8 \left(2 \Gamma ^3 \psi  (a \mu +\Gamma )+\left(\sqrt{a \Gamma ^7}+a \mu  \sqrt{a \Gamma ^5}\right) \dot \psi \right), \\
        &a^3 \mu  \phi ^{(4)}+a^2 \Gamma  \left(4 \mu  \ddot \phi -2 \psi ^{(4)}+\phi ^{(4)}\right)+4 a \Gamma ^2 \left(-2 B l^{(3)}+4 B \ddot \psi -2 \Delta  \dot l+4 \mu  \phi +\ddot \phi \right)+ \\
        &+4 \Delta  (a \Gamma )^{3/2} \ddot l+8 \sqrt{a \Gamma ^5} Q+16 \Gamma ^3 (\phi -\psi )=\Delta  \sqrt{a^5 \Gamma } l^{(4)}+8 \left(\sqrt{a \Gamma ^5}+\mu  (a \Gamma )^{3/2}\right) \dot \phi , \\
        &k \left(a^2 y^{(4)}+8 \Gamma ^2 y\right)=\frac{4 \Gamma ^{5/2} \dot Q}{\sqrt{a}}\>.
    \end{aligned}
\end{equation}
The resulting compressive force reads
\begin{multline}
    \mu^a=\frac{K}{q} + \frac{H^2 K q \left(H^4+2 \Delta  H^2-2\right)}{8 \Gamma ^2}+ q^2 \left(B-\frac{\Delta ^2 H^4}{8 \Gamma ^2}+\frac{1}{4}\right)+\frac{H^4 K q^3 \left(\Gamma ^2+H^4+10\right)}{32 \Gamma ^4}+ \\ +\frac{q^4 \left(4 B \Gamma ^2 H^2 \left(H^4+2 \Delta  H^2-2\right)+H^6 \left(\Gamma ^2-\Delta ^2\right)\right)}{32 \Gamma ^4}\>,
\end{multline}
where cell area $a$ has been expressed by cell height as $a=H^2/\Gamma$.

\subsection{No apico-basal tension polarity}
In an unsupported tissue with no apico-basal polarity (i.e., $\Delta=B=K=0$) the critical force
\begin{equation}
    \mu_{0, {\rm a}} = \frac{q^ 2}{4}
\end{equation}
does not depend on cell height and is calculated exactly the same as in the case where cell height is varied at a fixed cell cross-section area. Additionally we obtain the relation between the uniform height increase $\delta h$ and critical force $\mu$ as
\begin{equation}
    \delta h = \frac{\mu  \sqrt{a \Gamma }}{2 (\Gamma +\mu )} \approx \frac{\mu  \sqrt{a \Gamma }}{2 \Gamma }\>.
\end{equation}
Replacing $\sqrt{a\Gamma} = H$ and considering $\delta h/H=\epsilon$ we obtain the raltion for critical compressive strain
\begin{equation}
    \epsilon_0=\frac{\mu}{2\Gamma}\>,
\end{equation}
which is again unchanged by additional parameter $a$.

\subsection{Differential tension and basement membrane}
A tissue with no supporting stroma ($K=0, Q=0$) that has non-zero apico-basal polarity and basement membrane ($\Delta, B \neq 0$) exhibits a critical force
\begin{equation}
    \mu^{[\Delta, B]}_{\rm a} = q^2 \left(B-\frac{\Delta ^2 H^4}{8 \Gamma ^2}+\frac{1}{4}\right)+\frac{q^4 \left(4 B \Gamma ^2 H^2 \left(H^4+2 \Delta  H^2-2\right)+H^6 \left(\Gamma ^2-\Delta ^2\right)\right)}{32 \Gamma ^4}\>,
\end{equation}

In {\bf buckling regime} the first term on $\mu^{\Delta, B]}_a$ is sufficient to estimate critical force of a tissue with rest length $L$ as
\begin{equation}
    \mu_{\rm buck, a}^{\Delta, B} = \mu_{0, {\rm a}}\left(1+4B -\frac{\Delta^2 H^4}{2\Gamma^2}\right)\>.
\end{equation}
The onset of {\bf wrinkling regime} happens once $|\Delta|>\Delta_{\rm c, a}^{B} = \frac{\sqrt{2} \sqrt{(4 B+1) \Gamma ^2}}{H^2}$. Through the minimization of critical force
\begin{equation}
    \mu_{\rm wrink, a}^{\Delta, B} = q^2 \left(B-\frac{\Delta ^2 H^4}{8 \Gamma ^2}+\frac{1}{4}\right)+\frac{q^4 \left(4 B \Gamma ^2 H^2 \left(H^4+2 \Delta  H^2-2\right)+H^6 \left(\Gamma ^2-\Delta ^2\right)\right)}{32 \Gamma ^4}\>.
\end{equation}
we obtain the optimal wave-vector as
\begin{equation}
    q_{0, {\rm a}}^{[\Delta, B]} =\frac{\sqrt{2 \Delta ^2 H^4-4 (4 B+1) \Gamma ^2}}{\sqrt{4 B H^2 \left(H^4+2 \Delta  H^2-2\right)+H^6 \left(1-\frac{\Delta ^2}{\Gamma ^2}\right)}}
\end{equation}
and truncated to the first order around $\Delta_{\rm c, a}^{B}$ as
\begin{equation}
    q_{0, {\rm a}}^{[\Delta, B]} \approx \frac{2^{5/4}  \sqrt{-\Gamma  \left(4 \sqrt{2} B \Gamma -\sqrt{4 B+1} \Delta  H^2+\sqrt{2} \Gamma \right)} \sqrt{\Delta -\frac{\sqrt{8 B+2} \Gamma }{H^2}}}{\sqrt{8 \sqrt{2} B \sqrt{4 B+1} \Gamma +4 B H^4-16 B+H^4-2} \sqrt{\Delta  H^2-\sqrt{2} \sqrt{4 B+1} \Gamma }}
\end{equation}
\subsection{Stroma}
A tissue with no differential tension and basement membrane ($\Delta=B=0$), but supported by elastic stroma ($K \neq 0$) reaches instability at a critical force
\begin{equation}
    \mu_{\rm a}^{[K]}=\frac{\left(H^4-2\right) H^2 K q}{8 \Gamma ^2}+\frac{K}{q}+\frac{q^2}{4}\>.
\end{equation}
The optimal wave-vector obtained through minimization of $\mu_{\rm a}^{[K]}$ and truncated to the lowest order in $K$ around $K=0$, is
\begin{equation}
    q_{0, {\rm a}}^{[K]}=\left(2 K\right)^{1/3}\>.
\end{equation}